\documentclass[pra, aps, twocolumn, groupedaddress, showpacs, superscriptaddress]{revtex4-2}
\usepackage{amssymb,amsmath,amsthm,color,graphicx,times,graphicx}
\usepackage{hyperref}
\usepackage{subfigure}
\usepackage{times}
\usepackage{bbold}
\usepackage{color}
\usepackage{amsmath}
\usepackage{booktabs}
\usepackage{letltxmacro}
\usepackage{hyperref}
\hypersetup{
    colorlinks=true,
    linkcolor=black,    
    citecolor=black,    
    urlcolor=blue       
}

\providecommand{\openone}{\leavevmode\hbox{\small1\kern-4.3pt\normalsize1}}

\theoremstyle{plain}

\theoremstyle{definition}

\usepackage{orcidlink}

\begin{document}
\title{Unlocking thermodynamic multitasking: Exploring the functioning of two-qubit engines through coherence and entanglement}

\author{Hachem Tarif}\affiliation{LPHE-Modeling and Simulation, Faculty of Sciences, Mohammed V University in Rabat, Rabat, Morocco}
\author{Abdallah Slaoui \orcidlink{0000-0002-5284-3240}}\email{Corresponding author: abdallah.slaoui@fsr.um5.ac.ma}\affiliation{LPHE-Modeling and Simulation, Faculty of Sciences, Mohammed V University in Rabat, Rabat, Morocco} \affiliation{Centre of Physics and Mathematics, CPM, Faculty of Sciences, Mohammed V University in Rabat, Rabat, Morocco}
\author{Rachid Ahl Laamara \orcidlink{0000-0001-8410-9983}}\affiliation{LPHE-Modeling and Simulation, Faculty of Sciences, Mohammed V University in Rabat, Rabat, Morocco} \affiliation{Centre of Physics and Mathematics, CPM, Faculty of Sciences, Mohammed V University in Rabat, Rabat, Morocco}

\begin{abstract}

Recent studies have investigated the role of entanglement in the operation of a two-qubit system as a heat engine, showing that work can be extracted from a single heat bath without direct heat dissipation between the two-qubit system and the cold bath (2021 Phys. Rev. Lett, {\bf126} 120605). In this work, we explore the impact of operating the same two-qubit system model with two heat baths and direct dissipation to the environment by applying both a local and a global Markovian master equation. The addition of a second heat bath enables the system to operate in different modes depending on the initial quantum state. We examine the temporal behavior of concurrence entanglement and quantum coherence, analyzing their observable roles in transitions between various operational regimes. Additionally, we investigate the evolution of information flow throughout the working cycle of the two-qubit system, focusing on the influence of individual and collective decoherence on the system's efficiency and operational modes. We identify the optimal parameter regions for the engine and refrigerator modes to achieve maximum performance. Finally, we investigate the effect of coherence outside the system on its thermodynamic quantities.\par
\vspace{0.25cm}

\end{abstract}
\date{\today}

\maketitle
\section{Introduction}
In the pursuit of understanding the fundamental laws governing energy and matter, thermodynamics has long provided a cornerstone \cite{1,2}. However, as our exploration extends into the quantum realm, traditional thermodynamic principles transform, giving rise to quantum thermodynamics \cite{3,4,5,6,7,8,9}, where quantum coherence and entanglement play pivotal roles in shaping the behavior of microscopic systems. Quantum coherence \cite{10,11,12,13,14,15}, the delicate phase relationships among quantum states, and entanglement \cite{16,17}, the profound correlation between particles, emerge as key players in thermodynamic processes at the quantum level. This interdisciplinary frontier promises to deepen theoretical understanding and unlock revolutionary applications in energy, computation, and beyond.\par

Contemporary physics presents a range of systems that combine quantum mechanics and thermodynamics, such as quantum dots \cite{18,19}, superconducting circuits \cite{20,21}, trapped ions \cite{22,23}, and molecular machines \cite{24,25}. These systems exhibit remarkable phenomena like quantum phase transitions \cite{26}, quantum heat engines \cite{27,29,30,31,Kloc,Zhao}, and quantum information processing, challenging classical notions of thermodynamics and offering avenues for innovation. Quantum thermodynamics underpins the development of ultra-efficient nanoscale heat engines, quantum computers, and precision sensors \cite{32,33,34}, advancing fields such as renewable energy, materials science, and quantum communication \cite{35,36}.

At the core of quantum thermodynamics are quantum coherence and entanglement. Quantum coherence enables efficient energy transfer and information processing, while entanglement enhances the efficiency of quantum heat engines and allows for novel thermodynamic processes \cite{37,38,39}. These phenomena redefine thermodynamics at the microscopic level, offering unprecedented control over energy conversion and information manipulation in quantum systems. For example, coherence optimizes energy transfer pathways \cite{40,Palafox}. At the same time, entanglement facilitates cooperative interactions crucial for work extraction \cite{41}, as seen in systems like superconducting qubits or trapped ions, where coherence and entanglement enhance heat-to-work conversion efficiency. These examples underscore the vital role of quantum coherence and entanglement in advancing quantum thermodynamics.

In this paper, we explore the thermodynamic behavior of
a two-qubit quantum engine interacting with two heat baths. While traditional studies have extensively explored quantum coherence and entanglement in coupled qubit systems, extending to larger and more complex systems under local and global environmental interactions \cite{45,Nourmandipour}, our study seeks to transcend these established findings. We aim to bridge the microscopic quantum properties of this two-qubit system with macroscopic thermodynamic quantities such as efficiency and heat exchange with the environment. Furthermore, we provide a rigorous analytical derivation of the master equation in local and global descriptions, creating a robust framework that simplifies and facilitates future studies of two-qubit systems. Importantly, we establish a critical link between the system's mode of operation and foundational statistical principles by demonstrating that the probability distribution of the initial state significantly shapes the operational regime and thermodynamic behavior of this quantum system; this property is disregarded in many works \cite{50,51,Hu}.\par 

While the work in \cite{31} explores an alternative approach by eliminating one heat bath and relying on entanglement and the cold bath to operate their model, we instead emphasize the quantum significance of the heat bath in shaping the system's dynamics. The heat bath serves two purposes: it provides energy exchange and controls quantum coherence and information flow. We show this by looking at how the flow of information changes as the von Neumann entropy of the system changes. This perspective is pivotal in understanding the intricate interplay between quantum properties and thermodynamic behavior, offering a more comprehensive view of the heat bath's essential function in quantum system dynamics.\par

Notably, coherence and concurrence in this system do not necessarily act as direct thermodynamic resources for extracting additional work, as suggested in \cite{37,38,Streltsov2017}. Instead, their behavior emerges as a key factor in governing the transitions between the different operational modes of this two-qubit system. This insight drives our study to move beyond viewing quantum coherence and concurrence merely as resources; rather, we position them as critical indicators of the system's functional transitions. By doing so, we uncover their role in shaping the operational dynamics, offering a nuanced perspective on their contribution to the system's quantum thermodynamic behavior. While traditional analyses typically focus on the external factors affecting the system, we extend our investigation to explore the system's dynamics under both collective and individual decoherence effects. Unlike the work in \cite{Kumar}, which prioritizes power, we concentrate on heat exchange. This shift in perspective enables us to study the impact of decoherence on the heat exchange and efficiency of the two-qubit system. Our findings reveal an important result concerning the optimal spatial separation between the qubits, which enhances performance and promotes more favorable operating conditions.\par

Our motivation for this paper stems from a desire to enhance our understanding of the relationship between quantum information theory and the thermodynamic behavior of quantum systems. This work aims to contribute to advancing knowledge in this field. The paper is organized as follows: In Sec.\ref{sec.2}, we introduce the two-qubit system that serves as our model. We describe the system's dynamics using both the local master equation and the global master equation. In Sec.\ref{sec.3}, we analyze the engine's working cycle, modeled by the Otto cycle. We derive the heat and work quantities based on the two-qubit system's dynamics, we explore our results, examining how the engine's operating regime affects the system's quantum state. additionally, we investigate the impact of environmental coherence on the two-qubit system. Finally, Sec.\ref{sec.5} summarizes the key findings.

\section{Description of the model}\label{sec.2}
\begin{figure}[h!]
    \centering
    \includegraphics[scale=0.3]{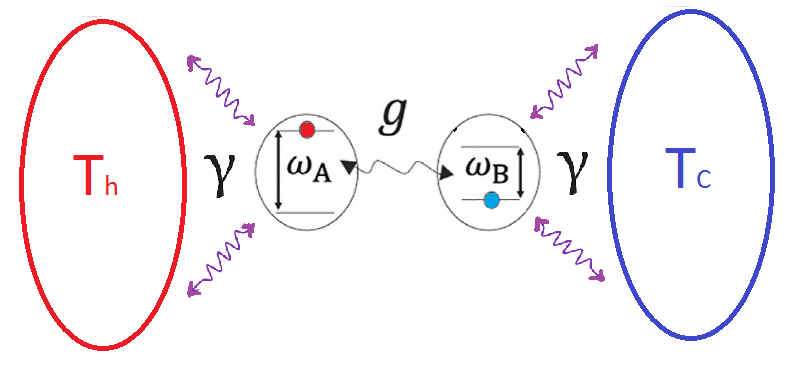}
    \caption{Schematic representation of the two-qubit heat engine. The two qubits are coupled to each other with a coupling strength $g$ as described by Eq.(\ref{eq.1}). The general form of the coupling of the qubit with each heat bath with a coupling strength $\gamma$, verifying the condition of $T_{h}>T_{c}$.}\label{fig:my_label}
\end{figure}

As sketched in Fig.(\ref{fig:my_label}), the quantum model considered is a two-qubit engine constructed using two-level qubits. The total Hamiltonian associated with this two-qubit heat engine, which comprises qubits $q_A$ and $q_B$, each linked to their respective baths: the hot bath $B_h$ and the cold bath $B_c$. These baths, indexed by $\alpha=\lbrace h,c\rbrace$, serve as the heat reservoirs. The total Hamiltonian of this two-qubit heat engine is expressed as 
\begin{equation}
H_{T}=\sum_{\alpha} H_{0,\alpha}+H_{I}+\sum_{\alpha}H_{D,\alpha}+\sum_{\alpha}H_{B,\alpha}, \label{H}
\end{equation}
where the free Hamiltonian of the two-qubit is expressed as
\begin{equation}
    H_{0,\alpha}=\omega_{\alpha} \sigma_{\alpha}^{\dagger}\sigma_{\alpha}.
\end{equation}
Each qubit, with transition frequencies $\omega_{A}$ and $\omega_{B}$, satisfies the condition $\omega_{B} < \omega_{A}$. We operate in natural units where $\hbar = k_{\beta} = 1$. The bosonic baths, which serve as thermal reservoirs for the system \cite{47,49}, are described as
\begin{equation}
H_{B,\beta}=\sum_{\lambda}\omega_{\lambda}b^{\dagger}_{\alpha,\lambda}b_{\alpha,\lambda}.\label{eq3}
\end{equation}
where the bath's field modes are indexed by $\lambda$, with $b_{\alpha,\lambda}$ and $b_{\alpha,\lambda}^{\dagger}$ representing their respective annihilation and creation operators, which satisfy the standard bosonic commutation relations.

Within the framework of local qubit-bath dissipative coupling, we consider terms of the form
\begin{equation}\label{eq4}
   H_{D,\alpha}= \sqrt{\gamma}(c_{z,\alpha}\sigma_{z,\alpha})\otimes \sum_{\lambda} \gamma_{\lambda}(b_{\alpha,\lambda}-b^{\dagger}_{\alpha,\lambda}),
\end{equation}
where $\sigma_{z,\alpha}$ is the Pauli z-operator representing the qubit's degree of freedom, $\gamma_{\lambda}$ are the coupling constants between the qubit and each oscillator mode $\lambda$, and $\gamma$ is the dissipation rate. We now need to define the interaction Hamiltonian between the two qubits \cite{31}, which, in our scenario, takes the form of
\begin{equation}\label{eq.1}
    H_{I}=\frac{g(t)}{2}(\sigma_{A}^{\dagger}\sigma_{B}+\sigma_{B}^{\dagger}\sigma_{A}),
\end{equation}
where $g$ is the interaction strength and $\sigma=|0\rangle \langle  1|$ is the lowering operator for the qubit $q_A$ and $q_B$. For the remaining calculations, we will use $H_{2qb}=\sum_{\beta} H_{0,\beta}+H_{I}$ to be the Hamiltonian of the two-qubit system. The DM coupling coefficient \( g \) characterizes the scenario in which the Dzyaloshinskii–Moriya (DM) anisotropy field is aligned along the \( z \)-axis\cite{Wang,Upadhyay,Mahmoudi}. In this configuration, the general DM interaction, given by \( \mathbf{D} \cdot (\mathbf{L} \times \mathbf{R}) \), simplifies to Eq.~(5), where \( \mathbf{D} = g \hat{z} \). This alignment emphasizes the role of \( g \) in determining the magnitude of the anisotropy field along the \( z \)-direction.\par
The Lindblad equation, a specific form of the master equation, describes the time evolution of an open quantum system interacting with its environment, leading to decoherence and dissipation. It provides a framework for modeling realistic quantum systems under external influences. The choice between local or global forms can affect calculations of work and heat dissipation. For our system (\ref{H}), its dynamics is conveniently described by

\begin{equation}
    \frac{d\rho_{S}}{dt}=-i[H_{2qb},\rho_{S}]+\mathcal{L}_{A}[\rho_{S}]+\mathcal{L}_{B}[\rho_{S}],
\end{equation}
where $\mathcal{L}_{B}$ and $\mathcal{L}_{A}$ are the Lindblad operators associated with the interactions of qubits $q_A$ and $q_B$, respectively. Here, we delve into two distinct formulations of the Lindblad superoperators as presented in the master equation: the Local Master Equation (LME) and the Global Master Equation (GME) \cite{44}:
\subsection{Local Master Equation}\label{sec7}
Here, we introduce the local Master equation, which describes a system's interaction with small parts of its environment at a time. This approach simplifies calculations by considering only the immediate environment's influence on the system's dynamics. However, in the context of work and heat dissipation calculations, a local approach might not capture the full complexity of energy exchanges between the system and its entire environment \cite{45}. It could overlook or approximate certain non-local interactions that are crucial for accurate assessments of work and heat dissipation. In this scenario, each qubit is connected to its environment through repeated interactions (see Fig.\ref{fig:my_label}). The two baths are described as a group of identical bosonic modes, with their behavior described by the Hamiltonian in equation (\ref{eq3}). The baths are prepared in thermal states, each with an inverse temperature of $\beta_{\alpha}$ where $\alpha \in \lbrace h,c\rbrace$, i.e.,
\begin{equation}\label{eq7}
   \rho_{B,\alpha}=\frac{e^{\beta_{\alpha}H_{B,\beta_{\alpha}}}}{tr(e^{\beta_{\alpha}H_{B,\beta_{\alpha}}})}.
\end{equation}
Following the construction of the well-known master equation based on the Born-Markov approximation, similar to the one used in \cite{46}, we can write the derivative of the density matrix in this form
\begin{equation}
    \frac{d\rho_{S}}{dt}=-i[H_{2qb},\rho_{S}]+\mathcal{L}^{(Loc)}_{\alpha}[\rho_{S}],
\end{equation}
defining the local thermal Lindblad operators acting on qubits as
\begin{equation}
    \mathcal{L}_{\alpha}[\rho_{S}]= \gamma_{\alpha}( n_{\alpha}\mathcal{D}[\sigma^{\dagger}_{\alpha}]+(n_{\alpha}+1)\mathcal{D}[\sigma_{\alpha}])
\end{equation}
where $\mathcal{D}(\sigma)=\sigma \rho_{S}\sigma^{\dagger}-\frac{1}{2}\lbrace \sigma^{\dagger} \sigma ,\rho_{S}\rbrace $ is the usual Lindblad dissipator, and
$n_{i}$ is the thermal occupation of the baths, given by $n^{\alpha}_{i}=1/(e^{\omega_{i}\beta_{\alpha}}-1)$ with $\gamma$ is the damping rate, $\sigma$ is the jump operator given of the form of the lowering and uppering operators. By examining the specific expression for the elements of the density matrix, we derive the following non-zero equations of motion:
\begin{align*}
    \mathcal{L}[\rho_{11}\left(t\right)]=\frac{d\rho_{11}\left(t\right)}{dt}&=-\left(\gamma_{A}^{+}+\gamma_{A}^{-}+\gamma_{B}^{+}\right)\rho_{11}\left(t\right)\notag\\&+\gamma_{B}^{+}\rho_{22}\left(t\right)+\gamma_{A}^{-}\rho_{33}\left(t\right),
\end{align*}
\begin{align*}
 \mathcal{L}[\rho_{22}\left(t\right)]= \frac{d\rho_{22}\left(t\right)}{dt}&=\gamma_{B}^{+}\rho_{11}\left(t\right)+\gamma_{A}^{-}\rho_{44}\left(t\right)\notag\\&+\left(\gamma_{A}^{-}+\gamma_{A}^{+}+\gamma_{B}^{-}\right)\rho_{22}\left(t\right),
\end{align*}
\begin{align}
 &\mathcal{L}[\rho_{33}\left(t\right)]= \frac{d\rho_{33}}{dt}=\gamma_{A}^{+}\rho_{11}\left(t\right)-\gamma_{B}^{+}\rho_{33}\left(t\right)-\gamma_{B}^{-}\rho_{44}\left(t\right),\notag\\&\mathcal{L}[\rho_{44}\left(t\right)]= \frac{d\rho_{44}\left(t\right)}{dt}=\gamma_{A}^{+}\rho_{22}\left(t\right)+\gamma_{B}^{+}\rho_{33}\left(t\right)-\gamma_{B}^{-}\rho_{44}\left(t\right),
\end{align}
at the same time, the off-diagonal density matrix elements, not coupled with the diagonal elements, obey the specified equations
\begin{equation*}
  \mathcal{L}[\rho_{23}]= \frac{d\rho_{23}}{dt}=\frac{1}{2}(\gamma_{A}^{+}+\gamma_{A}^{-}+\gamma_{B}^{+}+\gamma_{B}^{-})\rho_{23}(t),
\end{equation*}
\begin{equation*}
    \mathcal{L}[\rho_{12}]=\frac{d\rho_{12}}{dt}=-\frac{1}{2}(\gamma_{A}^{+}+\gamma_{A}^{-}+\gamma_{B}^{+}+\gamma_{B}^{-})\rho_{12}(t)+\gamma_{A}^{-}\rho_{34}(t),
\end{equation*}
\begin{equation*}
    \mathcal{L}[\rho_{13}]=\frac{d\rho_{13}}{dt}=-\frac{1}{2}(\gamma_{A}^{+}+\gamma_{A}^{-}+2\gamma_{B}^{+})\rho_{13}(t)+\gamma_{B}^{-}\rho_{24}(t),
\end{equation*}
\begin{equation*}
  \mathcal{L}[\rho_{14}]= \frac{d\rho_{14}}{dt}=\frac{1}{2}(\gamma_{A}^{+}+\gamma_{A}^{-}+\gamma_{B}^{+}+\gamma_{B}^{-})\rho_{14}(t),
\end{equation*}
\begin{equation*}
    \mathcal{L}[\rho_{24}]=\frac{d\rho_{24}}{dt}=-\frac{1}{2}(\gamma_{A}^{+}+\gamma_{A}^{-}+2\gamma_{B}^{-})\rho_{24}(t)+\gamma_{B}^{+}\rho_{13}(t),
\end{equation*}
\begin{equation*}
    \mathcal{L}[\rho_{34}]=\frac{d\rho_{34}}{dt}=-\frac{1}{2}(\gamma_{B}^{+}+\gamma_{B}^{-})\rho_{34}(t)+\gamma_{A}^{+}\rho_{12}(t).
\end{equation*}
where the quantities that appear in these elements are given by
\begin{align}
    &\gamma_{A}^{+}=(n_{A}+1)\gamma_{A},\hspace{1cm}\gamma_{A}^{-}=n_{A}\gamma_{A},\notag\\& \gamma_{B}^{+}=(n_{B}+1)\gamma_{B},\hspace{1cm}\gamma_{B}^{-}=n_{B}\gamma_{B}.
\end{align}
In the standard direct-product basis, defined by $|g_{1}g_{2}\rangle,|e_{1}g_{2}\rangle,|g_{1}e_{2}\rangle,|e_{1}e_{2}\rangle$, the density matrix elements for any prepared initial state are given by
\begin{align*}
     \rho_{11}&\left(t\right)=\left(\frac{\xi_{+}}{\eta_{+}+\eta_{-}}+\frac{\gamma^{+}_{B}+1}{\gamma^{+}_{B}}+\frac{\gamma^{+}_{A}e^{-t(\gamma^{+}_{B}+\gamma^{-}_{B})}}{\gamma^{-}_{B}+\gamma^{+}_{A}}\right)\rho_{11}\left(0\right)\notag\\& +\frac{\kappa}{\eta_{+}+\eta_{-}}\rho_{33}\left(0\right)
     + \left(\frac{\gamma^{+}_{B}+1}{\gamma^{+}_{B}}-\frac{\gamma^{+}_{B}e^{-t(\gamma^{+}_{B}+\gamma^{-}_{B})}}{\gamma^{-}_{B}+\gamma^{+}_{B}}\right)\rho_{22}\left(0\right),
     \end{align*}
\begin{align*}
     \rho_{22}&\left(t\right)=\left(\frac{\xi_{+}}{\eta_{+}+\eta_{-}}+\frac{\gamma^{+}_{B}}{\gamma^{+}_{B}+\gamma^{-}_{B}}+\frac{\gamma^{+}_{A}e^{-t(\gamma^{+}_{B}+\gamma^{-}_{B})}}{\gamma^{-}_{B}+\gamma^{+}_{A}} \right)\rho_{22}\left(0\right)\notag\\&+\frac{\kappa}{\eta_{+}+\eta_{-}} \rho_{44}\left(0\right)+  \left(\frac{\gamma^{+}_{B}}{\gamma^{+}_{B}+\gamma^{-}_{B}}-\frac{\gamma^{+}_{A}e^{-t(\gamma^{+}_{B}+\gamma^{-}_{B})}}{\gamma^{-}_{B}+\gamma^{+}_{A}}\right)\rho_{11}\left(0\right),
     \end{align*}
\begin{align*}
     \rho_{33}\left(t\right)&=\left(\frac{\xi_{-}}{\eta_{+}+\eta_{-}}+\frac{\gamma^{+}_{B}+1}{\gamma^{+}_{B}}+\frac{\gamma^{+}_{A}e^{-t(\gamma^{+}_{B}+\gamma^{-}_{B})}}{\gamma^{-}_{B}+\gamma^{+}_{A}}\right)\rho_{33}\left(0\right)\notag\\&+\frac{\chi}{\eta_{+}+\eta_{-}} \rho_{11}\left(0\right)+ \left(\frac{\gamma^{+}_{B}+1}{\gamma^{+}_{B}}-\frac{\gamma^{+}_{B}e^{-t(\gamma^{+}_{B}+\gamma^{-}_{B})}}{\gamma^{-}_{B}+\gamma^{+}_{B}}\right)\rho_{44}\left(0\right),
     \end{align*}
     
\begin{align*}
     \rho_{44}&\left(t\right)=\left(\frac{\xi_{-}}{\eta_{+}+\eta_{-}}+\frac{\gamma^{+}_{B}}{\gamma^{+}_{B}+\gamma^{-}_{B}}+\frac{\gamma^{+}_{A}e^{-t(\gamma^{+}_{B}+\gamma^{-}_{B})}}{\gamma^{-}_{B}+\gamma^{+}_{A}} \right)\rho_{44}\left(0\right)\notag\\&+\frac{\chi}{\eta_{+}+\eta_{-}} \rho_{22}\left(0\right)+\left(\frac{\gamma^{+}_{B}}{\gamma^{+}_{B}+\gamma^{-}_{B}}-\frac{\gamma^{+}_{A}e^{-t(\gamma^{+}_{B}+\gamma^{-}_{B})}}{\gamma^{-}_{B}+\gamma^{+}_{A}}\right)\rho_{33}\left(0\right),
 \end{align*}
     
\begin{align}
    \rho_{23}\left(t\right)=e^{-\frac{1}{2}\left(\gamma^{+}_{A}+\gamma^{-}_{A}+\gamma^{+}_{B}+\gamma^{-}_{B}\right)t}\rho_{23}\left(0\right).
\end{align}
with
\begin{align}
&\eta_{\pm}=\frac{\sqrt{{\gamma^{-}_{A}}^{2} + {\gamma^{+}_{A}}^{2} + 6\gamma^{-}_{A}\gamma^{-}_{A}}\pm(\gamma^{-}_{A} + \gamma^{+}_{A})}{2\gamma^{+}_{A}},\notag\\& \xi_{\pm} =\eta_{+}  e^{\eta_{\pm}\gamma^{+}_{A}  t} + \eta_{-}e^{\eta_{\mp}\gamma^{+}_{A}  t},\notag\\&\kappa =\eta_{+}\eta_{-}\left(e^{-\eta_{-} \gamma^{+}_{A}  t}- e^{\eta_{+}\gamma^{+}_{A} t}\right),\notag\\&\chi= e^{-\eta_{-} \gamma^{+}_{A}  t} - e^{\eta_{+}\gamma^{+}_{A} t}.
\end{align}

\subsection{Global Master Equation}\label{sec7}
We turn to the calculation of heat and work produced by the two-qubit engine, using the global master equation. This equation considers the entire system-environment interaction without making simplifying approximations, providing a more comprehensive view of how the system exchanges energy with its complete environment. Therefore, when analyzing work and heat dissipation with the global master equation, we can potentially capture a broader range of interactions and dynamics that contribute to energy exchanges between the system and its environment. This can lead to a more accurate assessment of the work done on the system and the heat dissipated to the environment. To construct this equation, we need to identify the jumping operators labeled $A_{\alpha}(\omega)$, where $\omega$ represents the transition rate between two quantum states.\par

To derive the global master equation, we need to follow the standard instructions found in many references \cite{47,49}. In this process, we determine the eigenvalues and their corresponding eigenvectors associated with the Hamiltonian of the system, denoted $H_{2qb}$; with $H_{2qb}|\epsilon\rangle=\epsilon|\epsilon\rangle$. The corresponding projection operators are given by
\begin{equation}
    \Pi(\epsilon)=|\epsilon\rangle\langle\epsilon|.
\end{equation}
The Hamiltonian of the bath remains unchanged from the previous subsection, as defined by equation (\ref{eq3}). To illustrate the procedure, we express the interaction Hamiltonian $H_{I}$ in the Schrödinger picture as follows 
\begin{equation}
    H_{I}=\sum_{\alpha}A_{\alpha}\otimes B^{\dagger}_{\alpha}+A^{\dagger}_{\alpha}\otimes B_{\alpha},
\end{equation}
where $A_{\alpha}=\sigma_{\alpha}$ and $B_{\alpha}=\sum_{k}g_{\alpha,k}b_{\alpha,k}$ are the system and bath operators, respectively \cite{47,48,49}. The Lindblad qubit operators are defined as
\begin{equation}
    A_{\alpha}(\omega)=\sum_{\omega}\Pi(\epsilon)A_{\alpha}\Pi(\epsilon'),
\end{equation}
where $\omega=\epsilon'-\epsilon$ represents the transition frequency, and the eigenoperators of the system Hamiltonian $A_{\alpha}(\omega)$ satisfy the following commutation relations 
\begin{equation}
    [H_{2qb},A_{\alpha}(\omega)]=-\omega A_{\alpha}(\omega),\hspace{0.4cm}
[H_{2qb},A^{\dagger}_{\alpha}(\omega)]=\omega A^{\dagger}_{\alpha}(\omega).
\end{equation}
In the interaction picture system-Bath coupling, the
Hamiltonian is written as
\begin{equation}
    H_{I}(t)=e^{iH_{2qb}t}H_{I}e^{-iH_{2qb}t}=\sum_{\alpha,\omega}e^{-i\omega t}A_{\alpha}(\omega)\otimes B_{\alpha}(t).
\end{equation}
with $B_{\alpha}(t)=\sum_{\lambda}g_{\lambda}(b_{\lambda}e^{-i\omega t}-b^{\dagger}_{\lambda}e^{i\omega t})$. To complete the construction of the reduced dynamics of the density matrix, as discussed in Refs.\cite{47,49} the spectral correlation tensor is
\begin{equation}\label{eq.19}
    \tau_{\alpha,\omega}=\omega^{3}e^{\beta \omega/2 }(\sinh(\beta \omega/2))^{-1}.
\end{equation}
For our system, the spectral correlation tensor is divided into four elements, each corresponding to a distinct transition frequency (see their analytical expressions in Appendix \ref{AppA}) . This led to the construction of the reduced dynamics of the density matrix using the nonzero eigen-operators, given as
\begin{equation}
    \frac{d\rho}{dt}=\sum_{\alpha,\omega}\gamma_{\alpha}\tau_{\alpha,\omega}( A_{\alpha}(\omega)\rho A_{\alpha}(\omega)^{\dagger}-\frac{1}{2}\lbrace A_{\alpha}(\omega)^{\dagger} A_{\alpha}(\omega) ,\rho \rbrace).
\end{equation}
The non-vanishing elements obtained from the previous equation are given by
\begin{align*}
    \mathcal{L}[\rho_{11}]=\frac{d\rho_{11}}{dt}&=\frac{1}{4}\left(\delta^{+}+\Omega^{+}\right)\left[\rho_{22}\left(t\right)+\rho_{33}\left(t\right)\right]\notag\\& +\frac{1}{4}\left(\delta^{-}+\Omega^{-}\right)\left[\rho_{23}\left(t\right)+ \rho_{32}\left(t\right)\right],
\end{align*}
\begin{align*}
  \mathcal{L}[\rho_{22}]= \frac{d\rho_{22}}{dt}&=\frac{1}{4}(\delta^{+}+\Omega^{+})\left[\rho_{44}\left(t\right)-\rho_{22}\left(t\right)\right]\notag\\&-\frac{1}{8}(\delta^{-}+\Omega^{-})\left[\rho_{23}\left(t\right)+\rho_{32}\left(t\right)\right],
\end{align*}
\begin{align*}
  \mathcal{L}[\rho_{33}]= \frac{d\rho_{33}}{dt}&=\frac{1}{4}(\delta^{+}+\Omega^{+})\left[\rho_{44}\left(t\right)-\rho_{33}\left(t\right)\right]\notag\\&-\frac{1}{8}(\delta^{-}+\Omega^{-})\left[\rho_{23}\left(t\right)+\rho_{32}\left(t\right)\right],
\end{align*}
\begin{align*}
  \mathcal{L}[\rho_{44}]= \frac{d\rho_{44}}{dt}=-\frac{1}{2}(\delta^{+}+\Omega^{+})\rho_{44}\left(t\right),
\end{align*}
\begin{align*}
  \mathcal{L}[\rho_{23}]= \frac{d\rho_{23}}{dt}&=-\frac{1}{4}(\delta^{+}+\Omega^{+})\rho_{23}(t)\notag\\&-\frac{1}{8}(\delta^{-}+\Omega^{-})\left[\rho_{22}(t)+\rho_{33}(t)+2\rho_{44}(t)\right],
\end{align*}
\begin{align*}
  \mathcal{L}[\rho_{12}]= \frac{d\rho_{12}}{dt}&=-\frac{1}{4}(\delta^{-}-\Omega^{-})\rho_{24}(t)-\frac{1}{4}(\delta^{+}-\Omega^{+})\rho_{34}(t)\notag\\&-\frac{1}{8}(\delta^{-}+\Omega^{+})\rho_{12}(t)-\frac{1}{8}(\delta^{+}-\Omega^{-})\rho_{13}(t),
\end{align*}
\begin{align*}
  \mathcal{L}[\rho_{13}]= \frac{d\rho_{13}}{dt}&=\frac{1}{4}(\delta^{+}-\Omega^{+})\rho_{24}(t)+\frac{1}{4}(\delta^{-}-\Omega^{-})\rho_{34}(t)\notag\\&-\frac{1}{8}(\delta^{-}+\Omega^{-})\rho_{12}(t)-\frac{1}{8}(\delta^{+}-\Omega^{+})\rho_{13}(t),
\end{align*}
\begin{align*}
  \mathcal{L}[\rho_{14}]= \frac{d\rho_{14}}{dt}&=-\frac{1}{4}(\delta^{+}+\Omega^{+})\rho_{14}(t),
\end{align*}
\begin{align*}
  \mathcal{L}[\rho_{24}]= \frac{d\rho_{24}}{dt}&=-\frac{3}{8}(\delta^{+}+\Omega^{+})\rho_{24}(t)-\frac{1}{8}(\delta^{-}+\Omega^{-})\rho_{34}(t),
\end{align*}
\begin{align}
  \mathcal{L}[\rho_{34}]= \frac{d\rho_{34}}{dt}&=-\frac{3}{8}(\delta^{+}+\Omega^{+})\rho_{34}(t)-\frac{1}{8}(\delta^{-}+\Omega^{-})\rho_{24}(t),
\end{align}
ensuring probability conservation $\sum_{i}\rho_{ii}=1$, $\rho_{ij}^{\dagger}=\rho_{ji}$ (with $i,j=1,2,3,4$), and the entries given by
\begin{align}
    &\delta^{+}=\gamma_h( \tau_{h,\omega_{+}}+\tau_{h,\omega_{-}}),\hspace{0.5cm}\delta^{-}=\gamma_h( \tau_{h,\omega_{+}}-\tau_{h,\omega_{-}}),\notag\\&\Omega^{+}=\gamma_c(\tau_{c,\omega_{+}}+\tau_{c,\omega_{-}}),\hspace{0.5cm}\Omega^{-}=\gamma_c(\tau_{c,\omega_{+}}-\tau_{c,\omega_{-}}).
\end{align}
To calculate the density matrix elements in the global master equation, as also cited in the local approach, we must specify the initial state. For an arbitrary two-qubit initial state, the density matrix elements are given by
\begin{align*}
\rho_{11}&\left(t\right)= \rho_{11}\left(0\right) + \rho_{22}\left(0\right) e^{-\frac{t}{2} \left(\Omega^{+} +\delta^{+}\right)}-\notag\\& 2\rho_{33}\left(0\right) e^{-\frac{t}{4}\left( \Omega^{-}+ \delta^{-} + \delta^{+}+\Omega^{+}\right)}+ 2\rho_{44}\left(0\right)e^{\frac{t}{4}\left(\Omega^{-}-\Omega^{+}+\delta^{-}-\delta^{+}\right)},
\end{align*}
\begin{align*}
\rho_{22}&\left(t\right)=-\rho_{22}\left(0\right)e^{-\frac{t}{2}(\Omega^{+}+ \delta^{+})}-\rho_{23}\left(0\right) e^{-\frac{t}{4}\left( \Omega^{+} + \delta^{+}\right)}+\notag\\&\rho_{33}\left(0\right) e^{-\frac{t}{4} \left(\Omega^{-}+\delta^{-} + \delta^{+} + \Omega^{+}\right)}- \rho_{44}\left(0\right) e^{\frac{t}{4}\left(\Omega^{-}-\Omega^{+}+\delta^{-} -\delta^{+}\right)},
\end{align*}
\begin{align*}
\rho_{33}\left(t\right)&= -\rho_{22}\left(0\right)e^{-\frac{t}{2}\left(\Omega^{+} +\delta^{+}\right)}+ \rho_{33}\left(0\right) e^{-\frac{t}{4}\left(\Omega^{-} +\delta^{-} +\delta^{+} +\Omega^{+}\right)}\notag\\&- \rho_{44}\left(0\right) e^{\frac{t}{4}\left(\Omega^{-}-\Omega^{+}+\delta^{-}-\delta^{+}\right)}+ \rho_{23}\left(0\right) e^{-\frac{t}{4} \left(\Omega^{+} +\delta^{+}\right)},
\end{align*}
\begin{align*}
\rho_{44}\left(t\right)= \rho_{22}\left(0\right)e^{-\frac{t}{2}\left( \Omega^{+} +\delta^{+}\right)},
\end{align*}
\begin{align}
\rho_{23}\left(t\right)= &2\rho_{33}\left(0\right) e^{-t \left(\frac{1}{4} \Omega^{-} + \frac{1}{4} \delta^{-} + \frac{1}{4} \delta^{+} + \frac{1}{4} \Omega^{+}\right)}\notag\\&-2 \rho_{44}\left(0\right) e^{t \left(\frac{1}{4} \Omega^{-} - \frac{1}{4} \Omega^{+} + \frac{1}{4} \delta^{-} - \frac{1}{4} \delta^{+}\right)}\notag\\&+\rho_{23}\left(0\right)e^{-\frac{t}{4} \left(\Omega^{+} +\delta^{+}\right)}.\label{eq23}
\end{align}
In the following, we consider these two approaches (local and global) and investigate the pivotal roles of coherence and entanglement in optimizing the performance of quantum thermal machines when our system is initially prepared in the state $|\phi (0) \rangle =\sqrt{p}|e_{1}g_{2} \rangle+\sqrt{1-p}|g_{1}e_{2} \rangle$. By analyzing a two-qubit engine under diverse conditions, we aim to uncover the intricate relationships among system parameters, efficiency, and quantum phenomena.

\section{Engine dynamics and functional modes in quantum states} \label{sec.3}
In this section, we delve into the potential functionalities of our setup as a thermal device. To achieve this, we employ both Lindblad and global master equations, along with heat flows and efficiency. In the first part (Sec.\ref{A}), we introduce the working cycle employed to operate the system—represented by the Otto cycle—and quantify the efficiency of the two-qubit system. Next (Sec.\ref{B}), we outline the possible modes through which the setup can function as a thermal machine, examining the behavior of coherence and entanglement in the system. Examining the thermodynamic behavior of the two-qubit system during the independent and collective decoherence effect in (Sec.\ref{D}).
Finally (Sec.\ref{C}), we address the influence of coherence within the heat bath on the thermodynamic behavior of the two-qubit system.

\subsection{Otto Cycle in a Two-Qubit System}\label{A}
\begin{figure}[h!]
    \centering
    \includegraphics[scale=0.25]{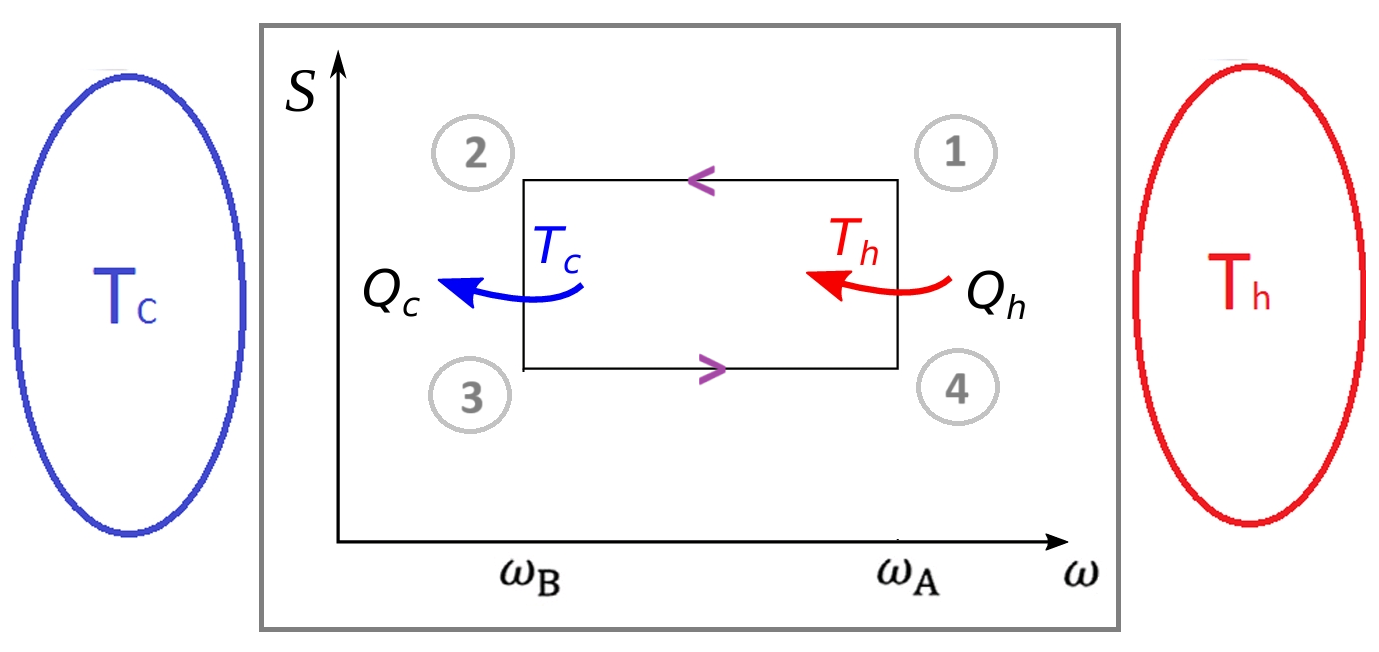}
     \caption{Schematic illustration of the Otto cycle in the entropy–transition frequency (S–$\omega$) diagram with hot and cold heat baths at temperatures $T_h$ and $T_c$, respectively.}\label{Otto_cycle}
\end{figure}

Quantum thermodynamic systems operate in versatile regimes, executing multitasking functions. Through the intricate interplay of coherence and entanglement, they efficiently convert energy while concurrently processing information. Harnessing these abilities is crucial for advancing quantum technology applications. In our model, the powering of the two-qubit engine relies on the utilization of the Otto cycle\cite{17,27}, renowned for its effectiveness, comprising four essential strokes as illustrated in the Figure(\ref{Otto_cycle}):\par
{\bf ($1$) Isentropic Compression:} The two-qubit engine undergoes decoupling from the hot bath, followed by tuning of its Hamiltonian, transitioning from $H_{A}\rightarrow H_{B}$. Throughout this stroke, work is applied to the system, inducing an elevation in its energy level, devoid of any heat exchange,
\begin{equation}
    \dot W=\frac{dW_{AB}}{dt}=tr(H_{2qb}(\mathcal{L}_{A}[\rho_{S}]+\mathcal{L}_{B}[\rho_{S}])).
\end{equation}
In this step, the system is in the initial state $|\phi (0) \rangle$. Additionally, we observe that the entropy remains constant, a consequence of the properties of Von Neumann entropy under unitary evolution.\par

{\bf ($2$) Cold Isochore:} Qubit $q_B$ is coupled to a cold bath to achieve thermal equilibrium, while its Hamiltonian remains fixed at $H_{B}$. This condition ensures that no work is done. Energy exchange occurs exclusively via heat transfer, represented as
\begin{equation}
    \dot Q_c=\frac{dQ_{c}}{dt}=tr(H_{2qb}\mathcal{L}_{B}[\rho_{S}]),
\end{equation}
where, as described in Ref.\cite{31}, the Lindblad operator changes between two initial states of the system. These two initial states are given by $|\phi (0) \rangle$ and $|\psi (0)\rangle =\cos(\theta)|e_{1}g_{2} \rangle-\sin(\theta) |g_{1}e_{2} \rangle $ in which the system is at his initial state, excited qubit $q_A$, and $q_B$ in the ground state with a probability of $p$ and excited qubit $q_B$, and $q_A$ in the ground state with a probability of ($1-p$). $\theta$ is given by $\arctan(g/\delta)$, $\delta$ is just the detuning of Rabi oscillator.\par

{\bf ($3$) Isentropic expansion:} Similar to the compression stroke, qubit $q_B$ is isolated from the cold bath, and the Hamiltonian transitions from $H_{B}\rightarrow H_{A}$. Again, under ideal conditions, the work performed is determined to be
\begin{equation}
    \dot W=\frac{dW_{BA}}{dt}=tr(H_{2qb}(\mathcal{L}_{A}[\rho_{S}]+\mathcal{L}_{B}[\rho_{S}])),
\end{equation}
and in this step, the system is in the initial state $|\psi (0)\rangle$.\par
{\bf ($4$) Hot isochore:} The cycle is completed by coupling qubit $q_A$ to the hot bath, allowing it to thermalize back to its initial state. The heat exchanged in this process is determined to be equal to
\begin{equation}
    \dot Q_h=\frac{dQ_{h}}{dt}=tr(H_{2qb}\mathcal{L}_{A}[\rho_{S}]),
\end{equation}
To incorporate information from \cite{17,27}, the evolution of the system under study is influenced by a Lindblad operator that varies between two distinct initial states, \( |\phi(0)\rangle \) and \( |\psi(0)\rangle \). This dynamic interplay between the two initial states provides a structured framework for investigating the system's evolution.\par
Conversely, Power as an important quantity in quantum thermodynamics, represents the rate of work performed by a quantum system, given by the time derivative of the work as
\begin{equation}
    P = \frac{dW}{dt}.
\end{equation}
By conserving the first law of thermodynamics, we can see that the total work extracted by the system when operating as a heat engine is the sum of the heat dissipated from the coupling between qubits and their corresponding baths, i.e., $W=Q_{c}+Q_{h}$. Consequently, the efficiency can be expressed as
\begin{equation}\label{eq.29}
    \eta=\frac{W}{Q_{h}}=1+\frac{Q_{c}}{Q_{h}},
\end{equation}
This efficiency is constrained by the Carnot limit, $\eta_{\rm Carnot} =1-T_{c}/T_{h}$,
reaching its maximum value at the lowest $T_{c}$ and highest $T_{h}$ temperatures.
It is important to note that the efficiency also depends on the specific cycle that powers the quantum thermal machine.\par

To evaluate the system's performance as a refrigerator, we employ the coefficient of performance (COP), defined as
\begin{equation}
    COP=\frac{Q_{c}}{W}=\frac{Q_{c}}{|Q_{c}+Q_{h}|}.
\end{equation}

\begin{figure}[h!]
    \centering
    \includegraphics[scale=0.3]{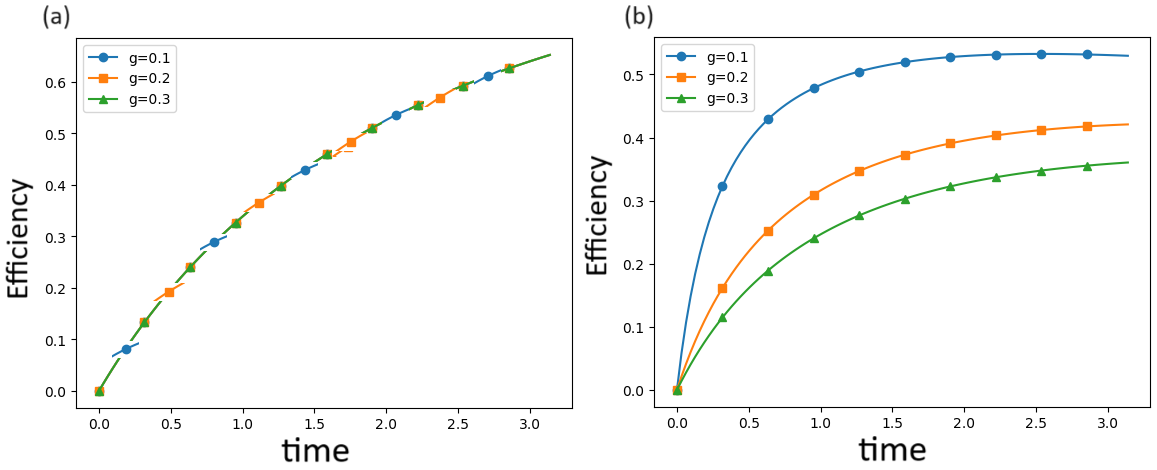}
    \caption{Efficiency, as defined in Eq.(\ref{eq.29}), plotted against time $t$ and coupling constant $g$ with $p=1$. (Panel $a$) Efficiency is calculated using the local master equation. (Panel $b$) Efficiency is calculated using the global master equation. Fixed parameters: $\omega_{A}=1$,$\omega_{B}=0.4$,$T_{c}=15$,$T_{h}=70$.}
    \label{2}
\end{figure}

In Fig.\ref{2}, we investigated the efficiency of a quantum heat engine using two different approaches: a local master equation (panel a) and a global master equation (panel b). Our analysis revealed contrasting behaviors in the efficiency of the heat engine concerning changes in the coupling strength between the two qubits. When employing the local master equation approach, we observed that the efficiency of the quantum heat engine remained unchanged, even when varying the coupling strength between the two qubits. This result suggests that the direct interaction between the two qubits, represented by the local environment, has limited influence on the overall efficiency of the heat engine. Consequently, the system's dynamics and performance appear to be largely insensitive to changes in the coupling strength within this framework. In contrast, our analysis using the global master equation approach yielded different outcomes. Here, we found that the efficiency of the quantum heat engine exhibited sensitivity to variations in the coupling strength between the two qubits. Specifically, as the coupling strength varied, so did the efficiency of the heat engine, indicating a significant influence of the collective behavior of the entire environment, including the interaction between the qubits, on the system's performance. This observation underscores the importance of considering correlations and non-local effects in capturing the full dynamics of the system-environment interaction.\par

Overall, our study highlights the importance of selecting an appropriate modeling framework based on the level of detail required and the specific characteristics of the system under investigation. While the local master equation offers a computationally efficient description of the system's dynamics, it may overlook important effects captured by the more comprehensive global master equation approach. Our results underscore the need to carefully consider modeling assumptions and their potential implications for interpreting experimental findings in quantum thermodynamics. In analyzing efficiency with the global master equation (Fig.\ref{2}b), we observed that increasing the coupling strength between the qubits led to a decrease in the efficiency of the quantum heat engine. This finding suggests that the qubit coupling plays a dissipative role, reducing the system's efficiency. One possible explanation for this behavior is that increased coupling strength enhances the interaction between the qubits and their environment, leading to stronger dissipative processes, such as energy loss or decoherence. Consequently, a greater portion of the input energy is lost to the environment, reducing the overall efficiency of the heat engine.
\begin{table}[htbp]
\centering
\caption{Conditions on Work and Heat Flow for different operations}
\label{ta.1}
\small 
\begin{tabular}{@{}lccc@{}}
\toprule
Operation & Heat flow $\dot{Q}_c$ & Heat flow $\dot{Q}_h$ & Work $\dot{W}$ (Q) \\ \midrule
Engine      & $\dot{Q}_c < 0$ & $\dot{Q}_h > 0$ & $\dot{W} < 0$ \\
Refrigerator & $\dot{Q}_c > 0$ & $\dot{Q}_h < 0$ & $\dot{W} > 0$ \\
Heat Pump   & $\dot{Q}_c < 0$ & $\dot{Q}_h > 0$ & $\dot{W} > 0$ \\
Accelerator & $\dot{Q}_c < 0$ & $\dot{Q}_h > 0$ & $\dot{W} > 0$ \\
Dissipator  & $\dot{Q}_c > 0$ & $\dot{Q}_h > 0$ & $\dot{W} = 0$ \\ 
\bottomrule
\end{tabular}
\end{table}

\begin{table}[htbp]
\centering
\caption{Conditions on Efficiency and Performance for different operations}
\label{ta.2}
\small 
\begin{tabular}{@{}lcc@{}}
\toprule
Operation & Efficiency and Performance Condition \\ \midrule
Engine      & $0 < \eta < 1-T_c/T_h$ \\
Refrigerator & $\text{COP}_{\text{R}} < T_c/(T_h-T_c)$ \\
Heat Pump   & $\text{COP}_{\text{HP}} < T_c/(T_h-T_c)$ \\
Accelerator & Not applicable (non-standard operation) \\
Dissipator  & Irreversible process \\ 
\bottomrule
\end{tabular}
\end{table}

\begin{figure}[h!]
    \centering
    \includegraphics[scale=0.3]{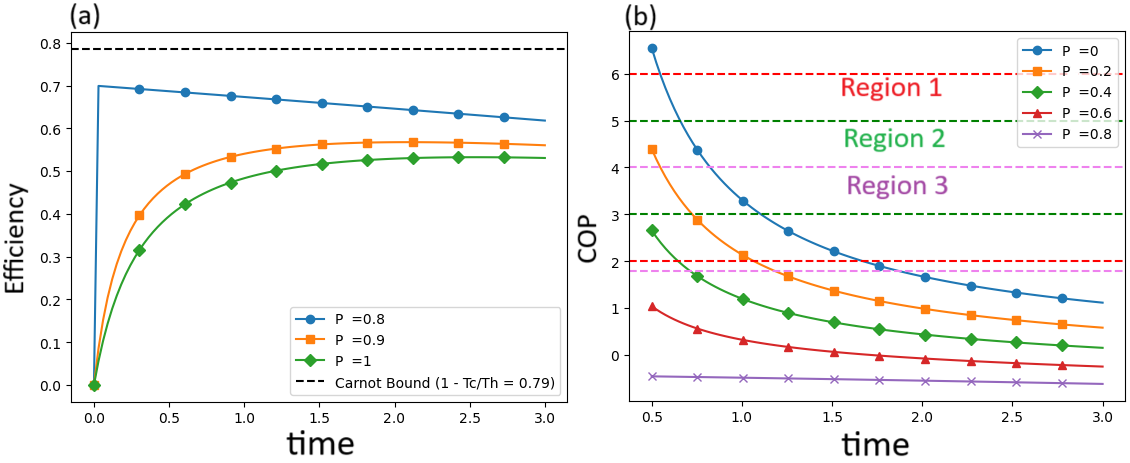}
    \caption{Efficiency (a) and coefficient of performance (b) for the two-qubit system, calculated using the global description of the dynamics, as a function of time and initial state probability $p$. Fixed parameters: $\omega_{A}=1$,$\omega_{B}=0.4$,$T_{c}=15$,$T_{h}=70$,$g=0.1$.}
    \label{3}
\end{figure}

The second result of this study focuses on the critical condition governing the selection of the system's initial state, which determines the engine's operational mode. Specifically, we show that the choice of the initial state dictates whether the system functions as a heat engine or a refrigerator. Through detailed analysis and numerical simulations (Fig.\ref{3}), we elucidate the precise conditions under which the system transitions between these two operational modes. This result underscores the significance of initial state selection in defining the thermodynamic behavior and operational characteristics of the system. Using the plots in Fig.(\ref{3}) and the additional insights provided in Appendix (\ref{AnxB}), along with the conditions from Table \ref{ta.1}, we can deduce that the probability associated with entering the initial state plays a pivotal role in determining the system's achievable operations. Our analysis reveals a distinct relationship between the initial state probability ($p$), time, and the system's operational mode. Specifically, we observe that for $p$ values ranging from $0.8$ to $1$ and time intervals from $0$ to $3$, the conditions required for the system to function as a heat engine are met. During this range, $Q_{c}<0$,$Q_{h}>0$ and $W<0$. Conversely, for \( p \) values in the range \( 0 < p \leq 0.4 \), and for time intervals that depend on each specific value of \( p \) as illustrated in Fig(\ref{3}), the system operates as a refrigerator. In this regime, the heat exchanges are characterized by \( Q_c > 0 \), \( Q_h < 0 \), and \( W > 0 \). For \( p \) values in the range \( 0.4 < p < 0.8 \), the system enters a non-functioning region for different time intervals. These findings highlight the nuanced interplay between initial state probability, operational time, and the resulting thermodynamic behavior of the system, offering valuable insights into its dynamic operation and performance characteristics.\par
It is significantly notable that the mode of operation also depends on time. We observe that for the engine operation, the optimal performance occurs at later times, whereas for the refrigerator mode, the best performance is observed at the beginning of the operation. This behavior can be explained by the nature of each machine. The absorption of heat from the cold bath, while operating as a refrigerator, primarily occurs at the start of the operation. In contrast, for the quantum engine to inject heat from the hot bath, it requires a certain amount of time to gain energetic momentum.\par
As shown in Fig.(\ref{3})a, the efficiency of the two-qubit engine approaches the Carnot limit when the initial population parameter, \( p \), is set to $0.8$. This proximity to the Carnot bound demonstrates the engine's optimized performance under this particular initial state configuration. Consequently, the system can extract a substantial amount of work while maintaining high thermodynamic efficiency, underscoring the effectiveness of the two-qubit setup in approaching theoretical efficiency limits typically reserved for idealized engines. 

The coefficient of performance is a critical measure of the efficiency of heat transfer systems, such as quantum refrigerators, quantum heat pumps, and quantum engines. This metric quantifies the ratio of useful work or heat transfer to the work input, and it varies depending on the type of cycle and the temperature conditions of the quantum system. The Carnot COP represents an idealized upper bound for efficiency, given by the ratio of the temperature difference between the hot and cold reservoirs to the temperature of the cold reservoir, 
\[
\text{COP}_{\text{Carnot}} = \frac{T_c}{T_h - T_c},
\]
where \( T_h \) and \( T_c \) are the absolute temperatures of the hot and cold reservoirs, respectively. This theoretical COP is achievable only in a reversible Carnot cycle, which serves as the benchmark for comparing real quantum systems. In practice, however, quantum systems experience losses due to non-idealities, leading to COP values typically lower than the Carnot limit. For example, in quantum refrigeration systems, COP values can range from $2$ to $6$ (Fig.\ref{3}b, Region $1$), influenced by the system's efficiency and the temperature gradient between the quantum system and its environment. Quantum air conditioners typically exhibit COP values between $2$ and $4$ under standard operating conditions (Fig.\ref{3}b, Region $3$), while quantum heat pumps in heating mode can achieve COP values between $3$ and $5$ (Fig.\ref{3}b, Region $2$), depending on the temperature differences. However, these values tend to decrease under extreme temperature conditions, where quantum thermal effects become more pronounced. Ultimately, the COP in quantum thermodynamic systems is determined by several factors, including system design, the quantum nature of the working medium, temperature differentials, and inherent thermodynamic inefficiencies as illustrated in Fig.(\ref{3})b.
\subsection{Temporal Behavior of Coherence and Entanglement} \label{B}

In an attempt to elucidate the observed results, we performed calculations of coherence and entanglement, recognizing their potential role in the quantum nature of our system. By quantifying coherence and entanglement, we aimed to investigate their influence on the system's thermodynamic behavior and operational modes. Our analysis sought to uncover possible correlations between these quantum properties and the system's tendency to operate either as a heat engine or a refrigerator. Through these calculations, we aimed to gain deeper insight into the underlying mechanisms driving the system's dynamic behavior and thermodynamic characteristics. \par 

{\bf Quantum Coherence:} QC, a key concept in quantum mechanics, refers to the ability of quantum systems to exist in superpositions of distinct states \cite{Streltsov2017,SlaouiB2023,Haddadi2024}. To quantify this, we use the intuitive $l_{1}$-norm coherence measure, defined as the sum of the absolute values of the off-diagonal elements of the quantum state in the computational basis. It is expressed as 
\begin{equation}
    \mathcal{C}_{l_{1}}=\sum_{ i \neq j }|\rho_{ij}|.
\end{equation}
Therefore, the $l_{1}$-norm coherence for our two-qubit engines using global approach (Eq.\ref{eq23}) can be expressed as follows
\begin{equation}
    \mathcal{C}_{l_{1}}=2e^{-(\delta^{+}+\Omega^{+})t/4} \sqrt{p (1 - p)}.\label{eq31}
\end{equation}

\begin{figure}[h!]
    \centering
    \includegraphics[scale=0.175]{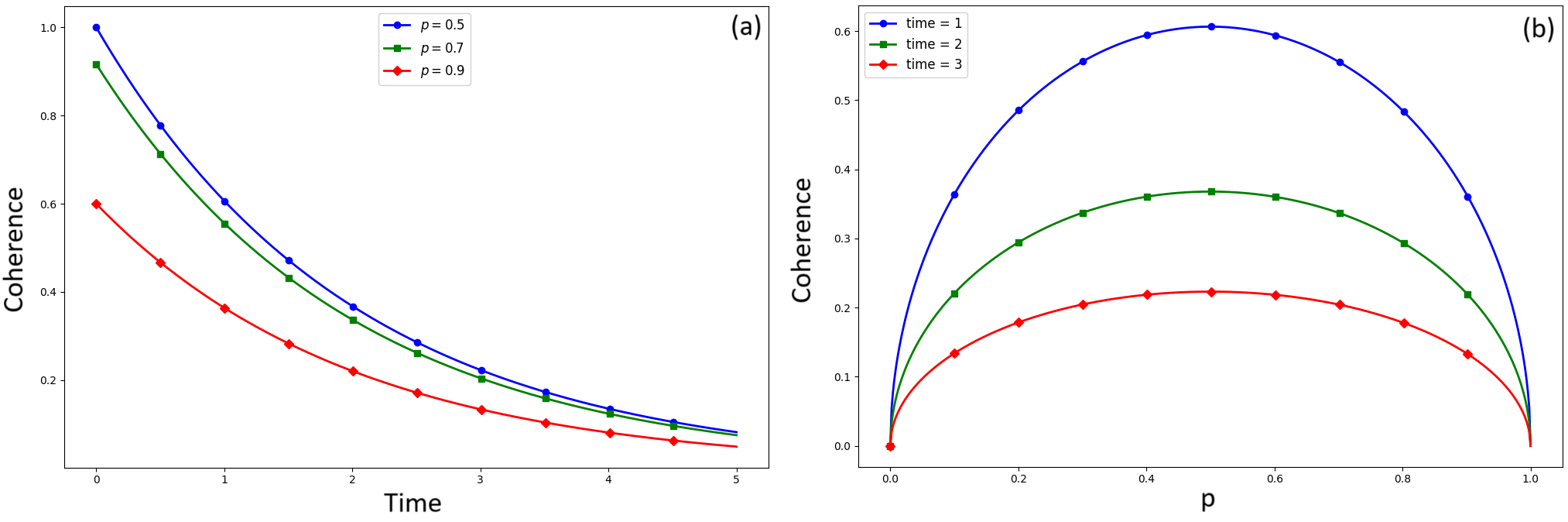}
     \caption{Coherence dynamics of a two-qubit system under a global description, (a) in the function of time for three values of the initial state probability $p$, (b) in the function of the initial state probability $p$ with three values of time with fixed parameters $\omega_{A}=1$, $\omega_{B}=0.4$, $T_{c}=15$, $T_{h}=70$, and $g=0.1$.}
    \label{4}
\end{figure}

In Fig.\ref{4}, based on Eq.(\ref{eq31}), we display the dynamics behavior of the $l_{1}$-norm coherence for fixed values of the initial population parameter $p$. We observe that the quantum coherence values reach their maximum at approximately \( p \approx 0.5 \), while they are initially at their peak at the beginning of the system's operation and then exhibit a slow decrease. Notably, at \( p \approx 0.5 \), the system undergoes a transition in its operational status, shifting from functioning as a heat engine to operating as a refrigerator, as illustrated in Fig.\ref{3}.


{\bf Information Flow:} The information flow from subsystem A to subsystem B provides valuable insight into the degree of communication between subsystems $q_A$ and $q_B$ in our quantum system. This information flow measure quantifies the amount of information shared between the two subsystems due to their common contribution to the system's operation. As the entropy of subsystem B increases, so does the degree of information passed between the subsystems, signifying a stronger correlation and mutual influence between their quantum states. Understanding the dynamics of entropy allows us to characterize the quantum information flow within the system and explore its implications for quantum information processing \cite{8}. The information flow for subsystems $q_A$ and $q_B$ can be calculated using the von Neumann entropy formula $S_{i} = -\text{Tr}(\rho_i \log_{2}(\rho_i))$,
where \(\text{Tr(.)}\) represents the trace operation and $\rho_B = \text{Tr}_A(\rho_{AB})$ and $\rho_A = \text{Tr}_B(\rho_{AB})$ are the reduced density matrices of the total system $\rho_{AB}$. Upon simplification, the entropies of subsystems $A$ and $B$ are determined to be
\begin{align}\label{35}
 S_{B}&=-e^{-(\delta^{+}+\Omega^{+})t/4}p \log_{2}(e^{-(\delta^{+}+\Omega^{+})t/4}p)-\notag\\&e^{-(\delta^{+}+\Omega^{+})t/4}\sqrt{p (1 - p)} log_{2}(e^{-(\delta^{+}+\Omega^{+})t/4}\sqrt{p (1 - p)},
\end{align}
and
\begin{align}\label{37}
    S_{A} = & - 
    \left( p e^{-\frac{t}{2}(\Omega^+ + \delta^+)} - (1-p)e^{-\frac{t}{4}(\Omega^- + \delta_- + \delta^+ + \Omega^+)} \right)  \nonumber \\
    &\log_{2} \left( p e^{-\frac{t}{2}(\Omega^+ + \delta^+)} - (1-p)e^{-\frac{t}{4}(\Omega^- + \delta_- + \delta^+ + \Omega^+)} \right)  \nonumber \\
    & + \left( (1-p)e^{-\frac{t}{4}(\Omega^- + \delta_- + \delta^+ + \Omega^+)} + e^{-\frac{t}{4}(\Omega^+ + \delta^+)}\left( p \right. \right. \nonumber \\
    & \left. \left. - \sqrt{p(1-p)}\right)  \right) \log_{2} \left( (1-p)e^{-\frac{t}{4}(\Omega^- + \delta_- + \delta^+ + \Omega^+)} \right. \nonumber \\
    &\left.   + \left(p  - \sqrt{p(1-p)}\right) e^{-\frac{t}{4}(\Omega^+ + \delta^+)} \right)  .
\end{align} 

\begin{figure}[h!]
    \centering
    \includegraphics[scale=0.31]{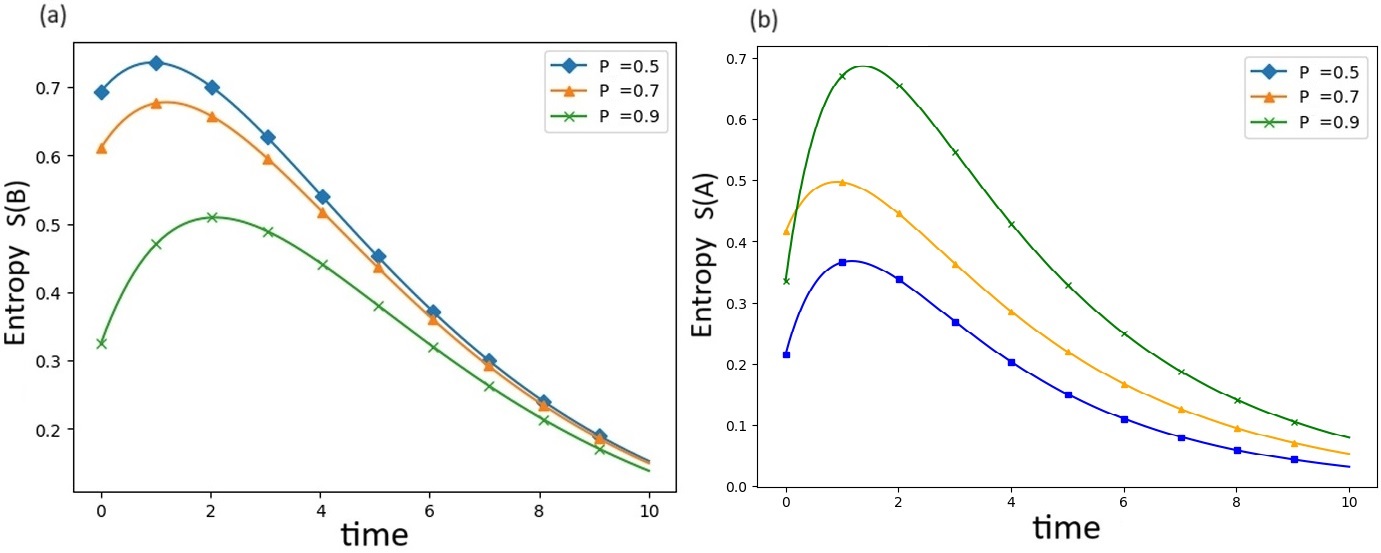}
    \caption{Entropy for subsystems $q_A$ (panel $a$) and $q_B$ (panel $b$) as a function of time and initial state probability $p$, calculated using the global description of the dynamics. Fixed parameters: $\omega_{A}=1$,$\omega_{B}=0.4$,$T_{c}=15$,$T_{h}=70$,$g=0.1$}
    \label{5}
\end{figure}

As shown in Fig.\ref{5}, the entropy dynamics display a pronounced peak at the system's midpoint, indicating a phase of maximal information flow and quantum coherence, during which the system’s state undergoes substantial transformations. For different values of $p$, the entropy behavior scales accordingly yet consistently reaches this peak, implying that interactions between states facilitate a transient phase of maximum uncertainty or disorder. As the evolution nears completion, the entropy rapidly decreases, suggesting a return to a more ordered or pure state as $t$ approaches the system’s final time. This decline in entropy reflects reduced information flow and coherence, suggesting a transition toward a stable configuration with minimized quantum information exchange. In our system, the two-qubit engine alternates its connection between the hot and
cold heat baths during the thermodynamic cycle. Consequently, the entropy of each qubit, which measures the information about their respective states, peaks around \(t=3\) for both qubits $q_A$ and $q_B$. During this interval, the engine’s performance and efficiency are maximized. At this stage, the heat baths thermalize the qubits, effectively erasing the information associated with  $S_{A}$ and $S_{B}$, as indicated by the decrease in entropy in Fig.~5 between $t=3$ and $t=10$. This entropy reduction corresponds to heat dissipation, returning the qubits to their respective initial states. This behavior highlights the role of von Neumann entropy in capturing the interplay between coherence and thermalization throughout the evolution.\par

{\bf Concurrence Entanglement:} Concurrence serves as a prominent metric for quantifying entanglement within two-qubit systems \cite{Wootters2001,Amghar2023,Haddadi2018,Rahman2000,Coffman2000}. For a specific category of density matrices known as X-states, characterized by their distinctive X-shaped structure, the calculation of concurrence can be streamlined by leveraging the eigenvalues of the spin-flipped density matrix. Mathematically, the concurrence (C) is expressed as 
\begin{equation}\label{36}
    {\cal C}(\rho_{AB}) = 2 \max(0, {\cal C}_{1}(\rho), {\cal C}_{2}(\rho)).
\end{equation}
Its analytical solution can be derived by calculating the values of ${\cal C}_{1}(\rho)$ and ${\cal C}_{2}(\rho)$, which are given by
\begin{align}
&{\cal C}_{1}(\rho) = \left| \rho_{14}(t) \right| - \sqrt{\rho_{22}(t) \rho_{33}(t)},\notag\\& {\cal C}_{2}(\rho) = \left| \rho_{23}(t) \right| - \sqrt{\rho_{11}(t) \rho_{44}(t)}.
\end{align}

\begin{figure}[h!]
    \centering
    \includegraphics[scale=0.175]{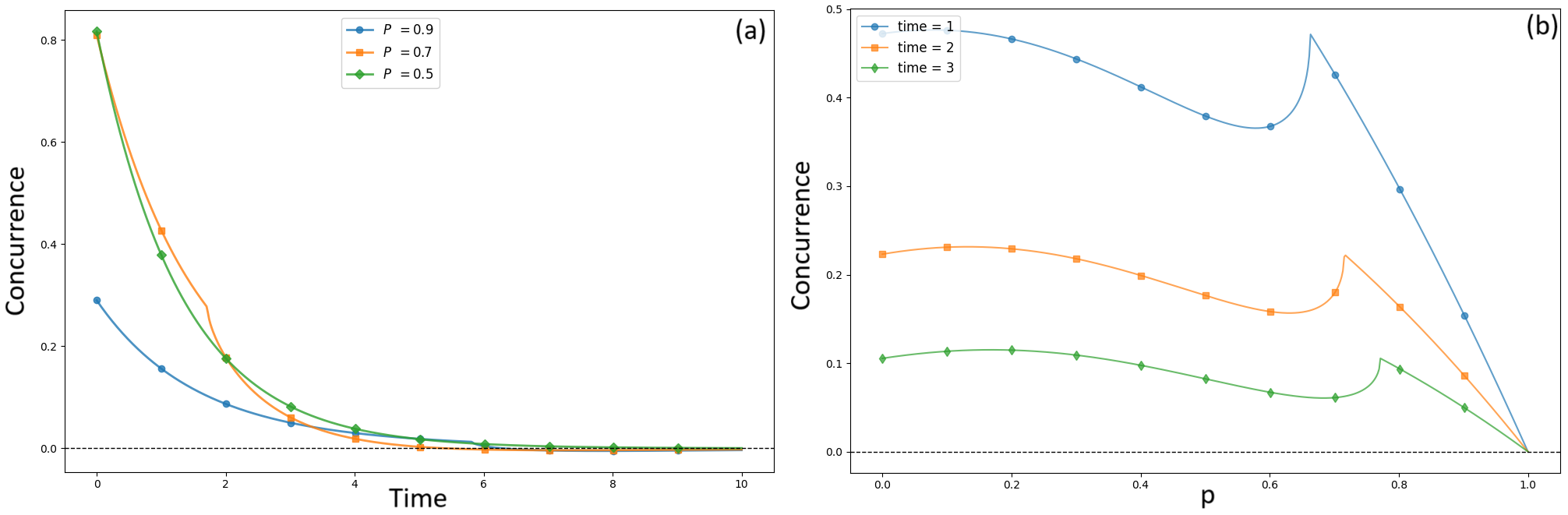}
    \caption{Dynamics of concurrence entanglement, calculated using the global master equation; Panel (a) for different values of initial state probability p during time evolution, Panel (b) the description of concurrence in the function of the initial state probability for three values of time . Fixed parameters: $\omega_{A}=1$, $\omega_{B}=0.4$, $T_{c}=15$, $T_{h}=70$, and $g=0.1$.}
    \label{6}
\end{figure}

The time evolution of concurrence is shown in Fig.~(\ref{6}.a), where its value decreases slowly over time. At \( p \approx 0.7 \), it exhibits an unusual peak for different time values Fig.~(\ref{6}.b). This peak occurs in the region of \( p \) where the system transitions in its operational status and also corresponds to the region where the system becomes non-functional. At this critical point, the system undergoes a significant transition in its thermodynamic behavior, switching from operating as a heat engine to a refrigerator. This critical value of $p$ marks a change in the working mode of the quantum system, where the entanglement is enhanced due to the strong correlations required for the refrigeration process. This behavior indicates that concurrence can serve as an indicator of mode-switching in quantum thermodynamic cycles. It reveals how entanglement plays a crucial role in defining the operational characteristics of quantum heat engines and refrigerators.
\begin{figure}[h!]
    \centering
    \includegraphics[scale=0.175]{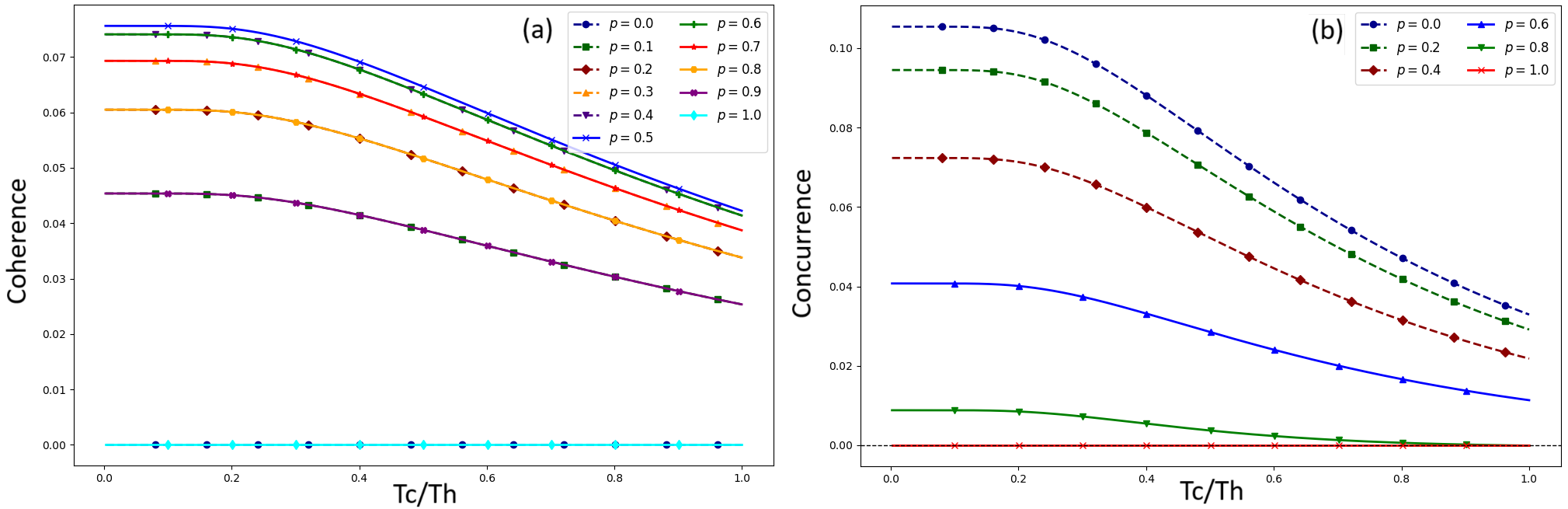}
     \caption{Dependence of Coherence (a) and Concurrence (b) on the Ratio \(T_c/T_h\) for Different Values of \(p\), with Fixed Parameters \(g = 0.1\) and \(t = 1\), in a Quantum System}
    \label{the Ratio}
\end{figure}

The two plots in Fig.\ref{the Ratio}($a$)-($b$) illustrate the behavior of coherence and concurrence, respectively, as functions of the temperature ratio \( T_c/T_h \), for various probabilities \( p \) associated with the quantum state. In panel ($a$), coherence demonstrates a general decreasing trend as \( T_c/T_h \) increases, reflecting the gradual loss of quantum superposition in the presence of thermal noise. This behavior aligns with the principle that higher temperatures induce greater thermal fluctuations, leading to enhanced decoherence effects. Importantly, the curves indicate that the degree of coherence at any given \( T_c/T_h \) strongly depends on \( p \), while values of \( p = 0.5 \) exhibit superior coherence preservation under thermal effects.

In panel ($b$), concurrence as given by equation(\ref{36}), a measure of entanglement, similarly exhibits a declining trend as \( T_c/T_h \) increases. The reduction in concurrence corresponds to the thermal disruption of entanglement between quantum subsystems, a key phenomenon in thermodynamic processes. Notably, concurrence diminishes more rapidly than coherence, suggesting that entanglement is more sensitive to temperature variations. The behavior of both metrics highlights their strong correlation with the thermodynamic entropy, as increasing temperature tends to increase disorder and reduce quantum correlations. These results underline the interplay between quantum information properties and thermodynamics, demonstrating how coherence and entanglement, fundamental resources in quantum technologies, degrade due to thermalization. By understanding these trends, we gain insights into optimizing quantum systems' performance under practical conditions, where minimizing decoherence and preserving entanglement is critical for robust quantum information processing and efficient quantum thermodynamic cycles.
\begin{figure}[h!]
    \centering
    \includegraphics[scale=0.225]{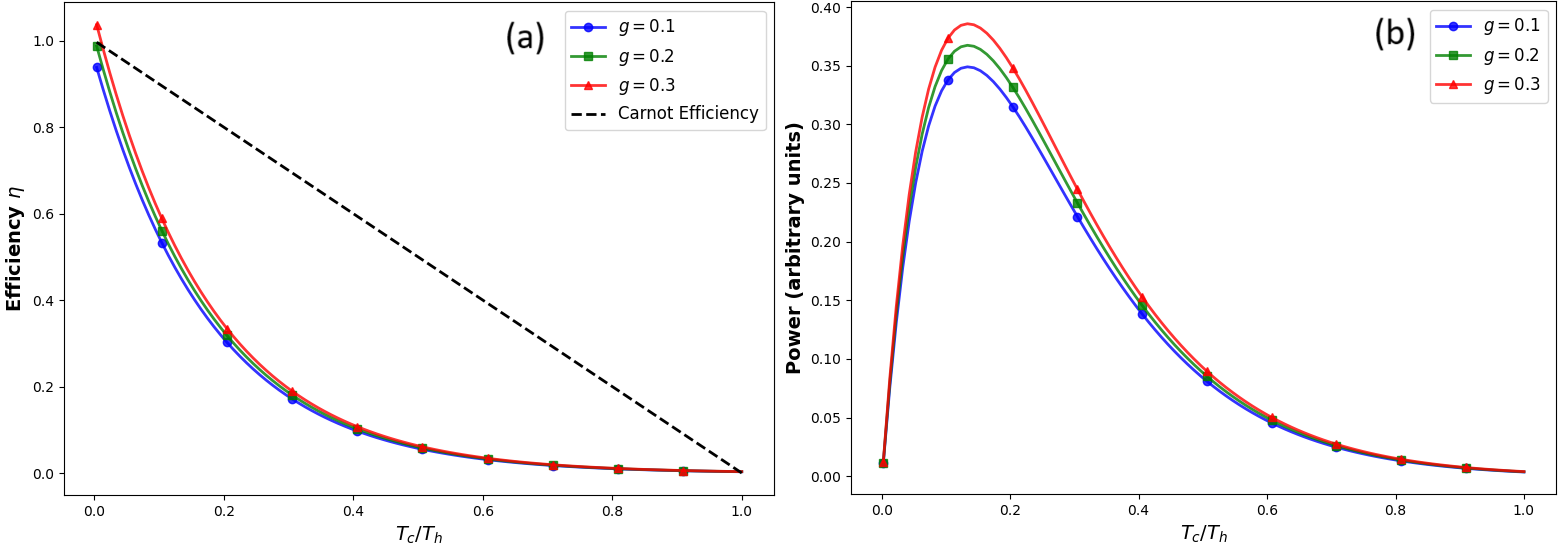}
    \caption{(a) Efficiency vs. Temperature Ratio for Three Coupling Strengths Between Two Qubits, (b) Power vs. Temperature Difference Ratio for Three Coupling Strengths}\label{Power}
\end{figure}

While incorporating all aspects of this two-qubit system in our study, it is important to emphasize the behavior of the engine’s efficiency under varying temperatures \(T_c\) of the cold bath and \(T_h\) of the hot bath. The efficiency, derived directly from Eq. (\ref{eq.29}), is plotted in Fig.~(\ref{Power}). As expected, the efficiency adheres to the Carnot bound and reaches its maximum when \(T_c \approx 0\) and \(T_h\) are significantly large. Furthermore, upon closer observation, we note that the efficiency becomes more dominant for larger values of \(g\). This behavior, which shows how important strong coupling is, only shows up when the effect of changing temperature is added to the expression of the spectral correlation tensor and looked at in the numerical plot, as shown in Eq. (\ref{eq.19}). Meanwhile, the power of a thermodynamic engine reflects how effectively it can perform work, making it a key metric for evaluating its performance. This power becomes even more intriguing in quantum systems due to the role of quantum coherence and interactions, which can shape the engine's behavior in unexpected ways. Interestingly, determining whether a device is operating as an engine or a refrigerator isn’t always straightforward when relying solely on the sign of the power. This ambiguity arises because work and energy flow definitions can differ depending on the context or framework used, leaving room for interpretation. In this context, our study highlights the power behavior of the two-qubit engine. We observe that its power peaks when the cold-to-hot temperature ratio is $T_c/T_h=0.2$, with significant values persisting for higher coupling strengths, as shown in Fig.\ref{Power}(b). This result points to the pivotal role of strong coupling in enhancing the engine’s performance. Moreover, the maximum power corresponds closely to the trends observed in the efficiency, showing how temperature gradients and coupling strength work together to shape the engine's operation. Although the power output alone can't definitively tell us how the system works, we are looking into it further by looking at how heat behaves, taking into account both its individual effects (when the qubits are separated) and its collective effects (when the qubits are coupled).

\subsection{Collective and Individual Effects on the Engine dynamics}\label{D}
The spectral correlation tensor \( \tau(\omega, r_{12}) \) encapsulates the interaction between two qubits and a shared thermal bath, with the qubits separated by a distance \( r_{12} \). It governs the decay rates \( \delta_+ \) and \( \Omega_+ \), which in turn influence heat exchange and efficiency in quantum systems. This tensor bridges the collective and individual decoherence regimes, depending on the spatial separation \( r_{12} \) and the wavevector \( k_0 \). The spectral correlation tensor is proportional to the spatial correlation function \( F(k_0 r_{12}) \), given by \cite{47,Kumar,Slaoui,Ficek}:
\begin{equation}
    \tau(\omega, r_{12}) \propto F(k_0 r_{12}),
\end{equation}
where
\begin{align}
F(k_0 r_{12}) =& \frac{3}{2} \bigg( 
 \left[ 1 - (\hat{\mu} \cdot \hat{r}_{12})^2 \right] 
\frac{\sin(k_0 r_{12})}{k_0 r_{12}}+\notag  \\
& \left[ 1 - 3 (\hat{\mu} \cdot \hat{r}_{12})^2 \right] 
\bigg[ 
\frac{\cos(k_0 r_{12})}{(k_0 r_{12})^2} - 
\frac{\sin(k_0 r_{12})}{(k_0 r_{12})^3} 
\bigg] 
\bigg),
\end{align}
and \( k_0 = \frac{2 \pi}{\lambda_0} \) is the wavevector, while \( \hat{\mu} \) and \( \hat{r}_{12} \) represent unit vectors of the dipole moment and the qubit separation, respectively.\par

The spatial correlation function \( F(k_0 r_{12}) \) plays a key role in determining whether two qubits interacting with a shared thermal bath experience collective or individual decoherence \cite{47,Kumar,Slaoui,Ficek}. When the qubits are close together, with \( k_0 r_{12} \ll 1 \), the bath modes interact strongly with both qubits at the same time. This leads to collective decoherence, enhancing decay rates and increasing qubit entanglement due to the shared environment. We can harness these collective effects to enhance coherence and facilitate quantum information processing. On the other hand, as the distance between the qubits increases, and \( k_0 r_{12} \gg 1 \), the correlation between the bath modes weakens. In this regime, the qubits decoherence independently because each interacts with a separate, uncorrelated part of the bath. The transition to individual decoherence signifies the loss of the collective effects. Understanding this shift from collective to independent behavior highlights how the distance \( r_{12} \) directly influences the qubits’ interactions with their environment, which is crucial for designing and optimizing quantum technologies.\par

The decay rates \( \delta_+ \) and \( \Omega_+ \) are directly proportional to \( \tau(\omega, r_{12}) \) as
\begin{equation}
\delta_+ = \gamma_h \tau(\omega_h, r_{12}),
\quad
\Omega_+ = \gamma_c  \tau(\omega_c, r_{12}),
\end{equation}
where \( \gamma_h \) and \( \gamma_c \) are system-bath coupling strengths in our system we set $ \gamma_h= \gamma_c= \gamma$. These rates govern the dynamics of heat exchange given by equations (B3) and (B4) in Appendix $B$. The transition from collective to individual effects significantly influences the heat exchange dynamics: At small separations \( r_{12} \), the heat absorbed (\( Q_h \)) and released (\( Q_c \)) are enhanced due to strong collective decoherence, where the shared thermal bath efficiently correlates the qubits and maximizes energy exchange. As \( r_{12} \) increases, these correlations weaken, leading to reduced \( Q_h \) and \( Q_c \) as the system transitions to independent decoherence, where each qubit interacts individually with the bath. This highlights the critical role of \( r_{12} \) in shaping heat dynamics, emphasizing the need to tailor qubit separation for optimal thermal and quantum performance (see Appendix \ref{AnxC}).
\begin{figure}[h!]
    \centering
    \includegraphics[scale=0.225]{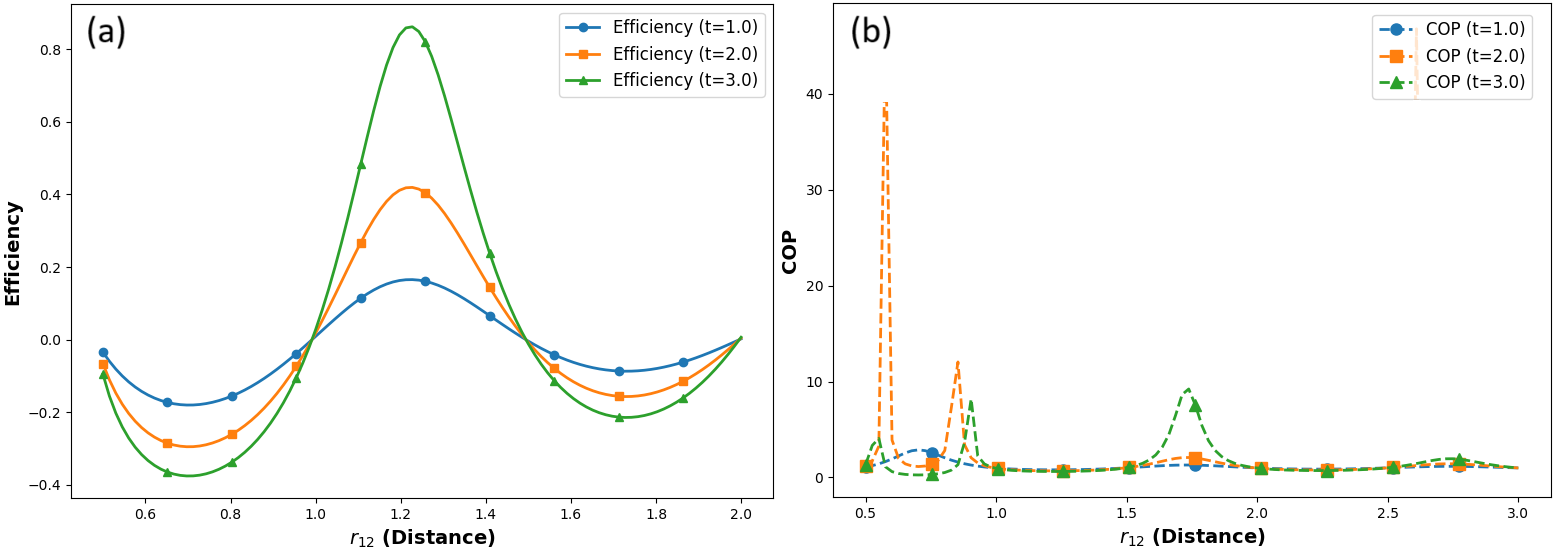}
    \caption{(a) Efficiency Behavior of the two-qubit engine in the function of the distance $r_{12}$ between the qubit $q_A$ and $q_B$, (b) the Coefficient of performance of the two-qubit fridge in the function of the distance $r_{12}$ between the qubit $q_A$ and $q_B$ for three values of time \{1, 2, 3\}. }\label{distance}
\end{figure}

The engine's efficiency, defined as equation (\ref{eq.29}), decreases with increasing \( r_{12} \), reflecting the transition from collective to individual decoherence. At small \( r_{12} \), where the spectral correlations are strong, efficiency is maximized due to enhanced heat absorption (\( Q_h \)) and minimized dissipation (\( Q_c \)). As \( r_{12} \) increases, the loss of these correlations reduces efficiency, highlighting the critical role of spatial separation in thermal performance. This behavior underscores the importance of the spectral correlation tensor in linking \( r_{12} \) to the decay rates \( \delta_+ \) and \( \Omega_+ \), which directly influence heat dynamics and efficiency. These findings provide valuable insights for designing quantum thermal devices and leveraging environment-mediated interactions to optimize quantum technologies. Respecting the working range of parameters that shape the behavior of the two-qubit system in Sec.(\ref{A}), we present Fig.~\ref{distance}, which demonstrates that the system exhibits an optimal distance for engine operation, with this distance varying across different time values. This contrasts with the refrigerator mode, where the distances for various time values differ. Furthermore, this study emphasizes the role of decoherence in shaping the thermodynamic behavior of the system. Specifically, we observe that collective decoherence (small $r_{12}$) and independent decoherence (large $r_{12}$) significantly influence heat flow into and out of the system, highlighting the impact of these decoherence mechanisms on the efficiency and performance of the system in both operational modes. Delving deeper into the analysis of Fig(\ref{distance}), we find that the engine mode Fig(\ref{distance},(a)) is maximum for a distinct value of $r_{12}\approx 1.2$ and experiences some nonfunctioning behavior in close range to this optimal distance due to the oscillatory behavior of the heat in the function of this distance. After this point and for larger $r_{12}$, the efficiency drops to zero. As for the fridge mode, the coefficient of performance exhibits some pics at different values of time and distance, while it seems at first glens that a larger value of (COP) in the function of $r_{12}$ is the best distance for functioning, this is not true in the thermodynamic framework. Respecting the regions mentioned in Sec.\ref{A}, we find that the best value of $r_{12} \approx 0.75$ for time=1 and in $r_{12} \approx \{ 1, 1.75\}$ for a time=3. As a point that combines the engine mode and the fridge mode, they both prefer the near-collective decoherence, as it is well numerically plotted in the Appendix \ref{AnxC}.\par

Following the purpose of this paper, we now turn to examining the effect of bath coherence on the system's functioning in the following section.

\subsection{Effect of bath coherence on system quantities} \label{C}
Here, we examine the effects of introducing a small amount of coherence into both system baths. Recall that these baths are modeled as bosonic baths, with their Hamiltonian described by Eq.(\ref{3}). The interaction Hamiltonian between the qubits and their respective baths is given in the form of Eq.(\ref{eq4}). Building on the insights from \cite{52,53}, we investigate the bath coherence generated by adding the off-diagonal term to Eq.(\ref{eq7}), represented by $\beta_{\alpha}$, with $\alpha \in \lbrace h,c\rbrace $
The density matrix \( \rho^{(i)}_R \) is given by
\begin{equation}
    \rho^{(i)}_R = \frac{e^{-\beta_i H^{(i)}_R}}{Z_i} + \lambda_i \sqrt{\tau} \xi,
\end{equation}
where the first term corresponds to the thermal Gibbs state at inverse temperature \( \beta_\alpha = \frac{1}{T_\alpha} \) (with Boltzmann’s constant set to \( k_B = 1 \)), and \( Z_\alpha = \text{Tr}[e^{-\beta_i H^{(i)}_{E,\beta}}] \) is the partition function. The second term, \( \xi \), is a traceless Hermitian operator with zero diagonal elements in the energy basis of \(H^{(i)}_{E,\beta}\). This term is defined as
\begin{equation}
    \xi =\mathbf{C}_{1} a + \mathbf{C}_{2} a^\dagger,
\end{equation}
represent the off-diagonal element responsible for coherence in the environment.
Here, \( a \) and \( a^\dagger \) are the annihilation and creation operators, respectively. The coefficients \( \mathbf{C}_{1} \) and \(\mathbf{C}_{2}\) are chosen to reflect the behavior of creation and annihilation operators in a coherent state. The parameter $\tau$ typically represents the coherence time, indicating the duration over which coherence is maintained \cite{52}. Finally, $\lambda_{i}$ quantifies the strength of the coherence injected into the bath.\par
To purely investigate the role of coherence in the work produced by a two-qubit heat engine, we consider the system under two types of interaction Hamiltonians: longitudinal coupling and transverse coupling \cite{54}. We then return to the interaction of Hamiltonian
\begin{equation}
 H_{D,\alpha}= \sqrt{\gamma}(c_{x_{\alpha}}\sigma_{x_{\alpha}}+c_{y_{\alpha}}\sigma_{y_{\alpha}}+c_{z_{\alpha}}\sigma_{z_{\alpha}})\otimes \sum_{\lambda} \gamma_{\lambda}(b_{\alpha,\lambda}-b^{\dagger}_{\alpha,\lambda}).
\end{equation} 
In the interaction Hamiltonian \(H_{D,\alpha}\), the coefficients \(c_{x_{\alpha}}, c_{y_{\alpha}},\) and \(c_{z_{\alpha}}\) determine the nature of the qubit-bath coupling. The terms \(c_{x_{\alpha}}\sigma_{x_{\alpha}}\) and \(c_{y_{\alpha}}\sigma_{y_{\alpha}}\) represent transverse coupling, which facilitates energy exchange between the qubit and the bath by inducing transitions between the qubit's \(|g\rangle\) and \(|e\rangle\) states. This type of coupling is crucial for processes that require direct energy transfer, such as the operation of quantum heat engines and refrigerators, where the system must absorb and emit energy quanta effectively. Conversely, the term \( c_{z_{\alpha}} \sigma_{z_{\alpha}} \) represents longitudinal coupling, which predominantly induces dephasing by influencing the population difference between the qubit states without generating state transitions. While longitudinal coupling contributes to dissipation, it does not drive the energy exchange essential for the thermodynamic cycles in quantum thermal machines. This limitation directly impacts the calculation of coherence-induced work. As described in \cite{52}, the work \( W_{i} \) is given by

\begin{equation}
    W_{i} := -i \lambda_i \beta_i \tau \left\langle \left[ G^{(i)}_{R}, H^{(i)}_{R} \right]  \right\rangle_{\xi^{(i)}},
\end{equation}
which arises from the unitary part of the evolution in the master equation \cite{53}. Here, we denote \( \langle \mathcal{O} \rangle_{\xi} = \text{Tr} \{ \mathcal{O}\xi \} \) as the trace over the coherent part of the initial state of the auxiliary units and \( G^{(i)}_R := \text{Tr}_i [ H^{(i)}_{D,\alpha} \rho_S ] \) is the effective Hamiltonian correction for each bath. In the case of transverse coupling, the effective Hamiltonian corrections for both baths—the cold bath and the hot bath—are zero. This implies that the interaction between the qubit and the baths does not alter the system's energy levels or induce transitions between the qubit states. Consequently, no energy exchange occurs between the qubit and the baths, meaning the qubit remains in its initial state without absorbing or emitting energy quanta. For quantum thermal machines, such as heat engines or refrigerators, this energy exchange is crucial. These machines operate by absorbing heat from the hot bath and releasing it to the cold bath, with work extracted during the process. Without Hamiltonian corrections, the machine cannot perform these essential thermodynamic processes. The zero effective Hamiltonian corrections suggest that the qubit is effectively decoupled from both the cold and hot baths in terms of energy transitions. This decoupling could result from the system's configuration, where interaction terms might cancel out due to symmetry or destructive interference, or from the absence of coherence in the bath, which is necessary for transverse coupling to induce effective energy exchange. While dissipation and relaxation processes still occur as part of the dissipative dynamics governed by the unitary evolution of the effective Hamiltonian, the qubit does not exchange energy with the baths in a way that would drive thermodynamic cycles.\par

As a result, quantum absorption refrigerators cannot function because they rely on the qubit to mediate heat transfer between the baths. Similarly, quantum heat engines cannot operate without the ability to absorb and release heat. Therefore, the qubit's lack of effective interaction with the baths renders the quantum thermal machine inoperative, necessitating a system configuration that supports coherence in the bath and ensures that the transverse coupling terms effectively induce the desired energy transitions. When transitioning to longitudinal coupling, we observe the same result, where \(c_{z_{\alpha}}\sigma_{z_{\alpha}}\) emerges as the only non-zero coupling coefficient. Consequently, the effective Hamiltonian corrections for both baths are zero, demonstrating that the injection of a small amount of coherence into the bath does not necessarily affect the thermodynamic quantities, as shown in \cite{52}. Instead, the impact depends on the inner configuration of the quantum system, which encompasses the state of each subsystem, described by basis states, the density matrix, or the wave function, as well as the interaction Hamiltonian that governs the dynamics.

\section{Concluding Remarks and Outlook}\label{sec.5}

In conclusion, this study provides valuable insights into the thermodynamic behavior of quantum systems, emphasizing the interplay between quantum coherence, entanglement, and system dynamics. By analyzing a two-qubit engine under various conditions, we illuminate the intricate relationship between system parameters, efficiency, and quantum phenomena. Our findings underscore the crucial role of coherence and entanglement in shaping the performance and functionality of quantum thermal machines by studying the behavior of concurrence and coherence in the function of the thermal parameter that shapes the operation of the two-qubit system. Heat baths are essential for initializing the system, erasing information (quantified by von Neumann entropy), and inducing heat dissipation, enabling the two-qubit engine to operate across different regimes as a function of the initial state probability. Unlike previous studies that relied on controlled operations, our setup leverages thermal interactions to drive the two-qubit system, paving the way for advancements in quantum technologies.\par
While the individual and collective effects shape the operation of this model, we demonstrated that there exists an optimal distance at which the two-qubit system can operate effectively. By focusing on the behavior of heat dissipation rather than power, this work highlights the significance of our findings and their contribution to the field. Looking ahead, future research directions may involve further investigating the impact of coherence and entanglement on the operation of quantum thermal machines and exploring strategies for enhancing their efficiency and stability. Additionally, leveraging recent developments in quantum control and optimization techniques could enable the realization of more robust and versatile quantum thermal machines. Remarkably, our theoretical framework provides a strong foundation for future exploration of larger spin-chain models, allowing for a deeper understanding of collective quantum phenomena and their role in quantum thermodynamic cycles. Extending this work to study non-Markovian effects in entangled spin systems could offer valuable insights into the dynamics of quantum coherence and entanglement, further bridging the gap between quantum information theory and thermodynamics. While we have demonstrated in the latest section that coherence in the environment does not necessarily affect the thermodynamic quantities, its impact depends on the configuration of the quantum machine, including the interaction Hamiltonian and the type of environment. Overall, this study contributes to the ongoing exploration of quantum thermodynamics and underscores the potential for harnessing quantum phenomena for technological innovation and scientific advancement.

\section*{Appendices}
In this section, we provide two appendices, A and B, which contain useful tools referenced throughout the main body of the paper. Appendix A presents explicit details on the derivation of the jump operators that enter in the construction of the global master equation. Appendix B outlines the behavior of the heat exchange in the thermodynamic cycle $Q_h$ and $Q_c$ in both the local and global master equation. Finally in Appendix C we will present the behavior of $Q_c$ and $Q_h$ in different values of the distance between the two-qubit in the two functioning modes, the engine, and the refrigerator. Showing the optimal value of $r_{12}$ for each functioning mode.

\appendix
\section{Derivation of the jump operators}\label{AppA}
We will proceed to systematically construct the jump operators of Markovian master equation governing the reduced state $\rho$ of the two-qubits engine, flowing the same procedure in \cite{44} and respecting the order of the states given by $|g_{1}g_{2}\rangle,|e_{1}g_{2}\rangle,|g_{1}e_{2}\rangle,|e_{1}e_{2}\rangle$
the Hamiltonian of the system in the natural units is given as
\begin{equation*}
H_{2qb} = 
\begin{pmatrix}
0 & 0 & 0 & 0 \\
0 & \omega_{A} & g/2 & 0\\
0 & g/2 & \omega_{B} & 0\\
0 & 0 & 0 & \omega_{A}+\omega_{B}
\end{pmatrix},
\end{equation*}
where the eigenvalue that construct the transition rate $\omega$ is 
\begin{align*}
  \lbrace 0,\omega_{B}+\omega_{A},&\frac{(\omega_{B}+\omega_{A})+\sqrt{(\omega_{A}-\omega_{B})^{2}+g^{2}}}{2},\notag\\&\frac{(\omega_{B}+\omega_{A})-\sqrt{(\omega_{A}-\omega_{B})^{2}+g^{2}}}{2} \rbrace.
\end{align*}
As well as their associated eigenvectors
\begin{align}
  &|1\rangle=|g_{1}g_{2}\rangle,\hspace{0.5cm}|3\rangle=\frac{1}{\sqrt{2}}(|e_{1}g_{2}\rangle+|g_{1}e_{2}\rangle), \notag\\&|2\rangle=|e_{1}e_{2}\rangle,\hspace{0.5cm}|4\rangle=\frac{1}{\sqrt{2}}(|e_{1}g_{2}\rangle-|g_{1}e_{2}\rangle).
\end{align}
We can now easily derive the jump operators using (Eq. 10) written as follow:

\begin{equation}
A_{h}(\omega_{\pm}) = 
\begin{pmatrix}
0 & 1/2 & \pm1/2 & 0 \\
0 & 0 & 0 & \mp1/2\\
0 & 0 & 0 & 1/2\\
0 & 0 & 0 & 0
\end{pmatrix},
\end{equation}
and
\begin{equation}
A_{c}(\omega_{\pm}) = 
\begin{pmatrix}
0 & \pm1/2 & 1/2 & 0 \\
0 & 0 & 0 & 1/2\\
0 & 0 & 0 & \mp1/2\\
0 & 0 & 0 & 0
\end{pmatrix},
\end{equation}
in which 
\begin{align}
 &\omega_{+}=\frac{(\omega_{B}+\omega_{A})+\sqrt{(\omega_{A}-\omega_{B})^{2}+g^{2}}}{2},\notag\\& \omega_{-}=\frac{(\omega_{B}+\omega_{A})-\sqrt{(\omega_{A}-\omega_{B})^{2}+g^{2}}}{2},  
\end{align}
is the only transition rate with non-vanishing Lindblad operators.
\section{Dynamic behavior of $Q_{h}$ and $ Q_{c}$ in the local and global master equation}\label{AnxB}
In this appendix, we will provide the exact expressions for the heat flow and analyze their dynamic behavior within the framework of both the local and global master equations, specifically for $p=1$. We begin by presenting the results for the local description of the system, where we have derived the following expressions
\begin{align}\label{B1}
    Q_{hL}&= g\cos\theta\sin\theta\exp\left(\frac{(\gamma^{+}_{A} + \gamma^{-}_{A}) t}{2}\right) +\notag
    \\&\omega_{A} \left(\frac{\xi_{+}}{(\eta_{+}+ \eta_{-})}\right)(\cos^{2}\theta - 1) +\omega_{B}\sin^{2}\theta\left(\frac{\xi_{-}}{(\eta_{+} + \eta_{-})}\right)  \notag\\& +(\omega_{A}+\omega_{B}) \left(\frac{\kappa}{(\eta_{+} + \eta_{-})}\right) \sin^{2}\theta,
\end{align}
and 
\begin{align}\label{B2}
Q_{cL}&= -\cos\theta\sin\theta g\exp\left(\frac{(\gamma^{+}_{B} + \gamma^{-}_{B}) t}{2}\right)\notag\\& + \left(\frac{\gamma^{+}_{B}+\gamma^{-}_{B}  \exp(t  (\gamma^{+}_{B} + \gamma^{-}_{B}))}{(\gamma^{+}_{B} + \gamma^{-}_{B})}\right) \omega_{A} \left(1 - \cos^{2}\theta\right) \notag\\&- \omega_{B}  \left(\frac{\gamma^{-}_{B}}{(\gamma^{+}_{B} + \gamma^{-}_{B})} + \frac{\exp(t  (\gamma^{+}_{B} + \gamma^{-}_{B}))}{((\gamma^{+}_{B} + \gamma^{-}_{B}) + 1)}\right)  \sin^{2}\theta +\notag\\& \frac{\gamma^{-}_{B}\left(1 - \cos^{2}\theta\right)}{(\gamma^{+}_{B} + \gamma^{-}_{B})}\left(1-\exp(t  (\gamma^{+}_{B} + \gamma^{-}_{B}))\right)  (\omega_{A}+\omega_{B}).
\end{align}
 In a similar manner to the local description, we find the corresponding expressions for the heat flows $Q_{h}$ and $Q_{c}$ in the global description. This allows us to examine the heat exchange dynamics from a more comprehensive perspective on a global scale. Their analytical expressions are given by
\begin{align}\label{B3}
 Q_{h}=&g\left(\cos\theta\sin\theta \exp\left(-\frac{\delta^{+}t}{ 2}\right) - \sqrt{p  (1-p)}\right) \notag\\&+\omega_{A}  \left(\exp\left(-\frac{\delta^{+}t}{ 2}\right)  (\cos^{2}\theta) - p\right) \notag\\&+ \omega_{B}  \left(\exp\left(-\frac{\delta^{+}t}{2}\right)  (\sin^{2}\theta) - (1 - p)\right),   
\end{align}
and
\begin{align}\label{B4}
Q_{c}=&-g  \left(\cos\theta\sin\theta \exp\left(-\frac{\Omega^{+}t}{ 2}\right) - \sqrt{p (1 - p)}\right) \notag\\&+ \omega_{A}  \left(p - \exp\left(-\frac{\Omega^{+}t}{d \cdot 2}\right) (\cos^{2}\theta)\right) -\notag\\& \omega_{B} \left(\exp\left(-\frac{\Omega^{+}t}{2}\right)  (\sin^{2}\theta) - (1 - p)\right).    
\end{align}

Figures \ref{Fig7}(panels $a$ and $d$) and \ref{Fig7}(panels $b$ and $e$) respectively depict the numerical plots of heat absorbed by the system (\ref{B1}) and released to the cold bath (\ref{B2}) in the local master equation description, and the heat absorbed by the system (\ref{B3}) and released to the cold bath (\ref{B4}) in the global master equation description. To elucidate the effect of $p$ on $Q_h$ and $Q_c$ , plotted in Fig.\ref{Fig7}(panels $c$ and $f$), we examine the heat flow as a function of time for various $p$ values under both local and global density matrix dynamics. We observe that for $p=1$, increasing the coupling g from $0.1$ to $0.3$ leads to greater heat absorption by the system $Q_h$ (panels a and b) and smaller heat release to the cold bath $Q_c$ (panels d and e), regardless of the dynamic description. This underscores the significance of coupling strength g in energy flow. For a fixed $g=0.1$ and varying initial state probability $p$ between $0$ and $1$, the system transitions from a quantum fridge to a heat engine at p=1 (panels $c$ and $f$), as confirmed by Table (\ref{ta.1}). This highlights the crucial role of initial system state in determining its operational mode.

\section{The behavior of efficiency and COP in the Collective and Individual decoherence}\label{AnxC}
Respecting the conditions of functioning mentioned in section(3) for each of the engine and fridge modes, the Fig.(\ref{Fig12}) is constructed. As is well illustrated in Fig.(\ref{Fig12}), the quantum system presented as the two-qubit has a near preference for the collective decoherence mode. While for large values of \( r_{12} \), where the individual decoherence dominates, the values of both the efficiency and the coefficient of performance (COP) decrease notably, as shown in Figure (\ref{Fig12}). Although that is true, this part highlights other conditional elements, where we see clearly that even in the near-collective decoherence regime, there are specific values of \( r_{12} \) in which the two-qubit system does not function as a thermal machine.\par
 
To further explore this intriguing behavior, we turn to the numerical plots of the heat exchange, \( Q_c \) and \( Q_h \), presented in Fig.(\ref{Fig12}, panels (a) and (d)). These plots reveal a distinct oscillatory pattern in the heat values across both the engine mode (panel (a)) and the refrigerator mode (panel (d)). This oscillatory behavior, rooted in quantum coherence and interference effects, underscores the complex interplay between decoherence mechanisms and thermodynamic performance in quantum systems. Understanding these oscillations is key to unraveling the nuanced conditions under which the system transitions between functional and non-functional thermal machine states. This oscillation could stem from the inherent nature of quantum mechanics. Energy's inexplicable oscillatory behavior often originates from quantum coherence and interference effects, where the wave-like nature of particles induces energy fluctuations between states, defying classical intuition.

\section*{ACKNOWLEDGMENTS}
H.T. acknowledges the financial support of the National Center for Scientific and Technical Research (CNRST) through the “PhD-Associate Scholarship-PASS” program. The authors acknowledge the LPHE-MS, FSR for the technical support.\par 

{\bf Declaration of competing interest:} The authors declare that they have no known competing financial interests or personal relationships that could have appeared to influence the work reported in this paper.\\

{\bf Data availability:} No data was used for the research described in the article.
\begin{widetext}

 		\begin{figure}[hbtp]
 			{{\begin{minipage}[b]{.33\linewidth}
 						\centering
 						\includegraphics[scale=0.4]{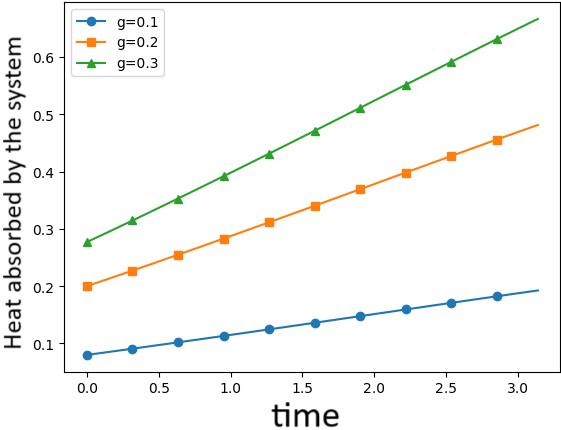} \vfill $\left(a\right)$
 					\end{minipage} \hfill
 					\begin{minipage}[b]{.33\linewidth}
 						\centering
 						\includegraphics[scale=0.4]{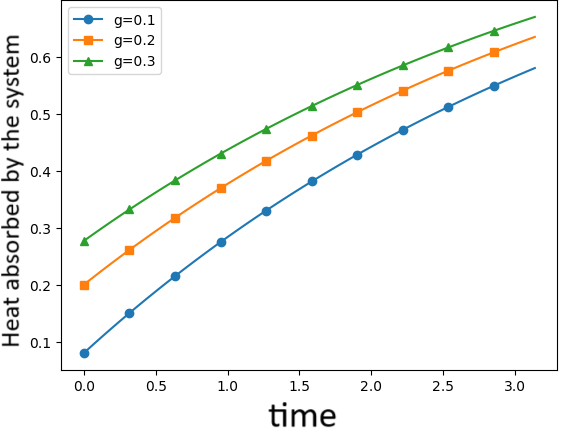} \vfill  $\left(b\right)$
 					\end{minipage} \hfill
 					\begin{minipage}[b]{.33\linewidth}
 						\centering
 						\includegraphics[scale=0.4]{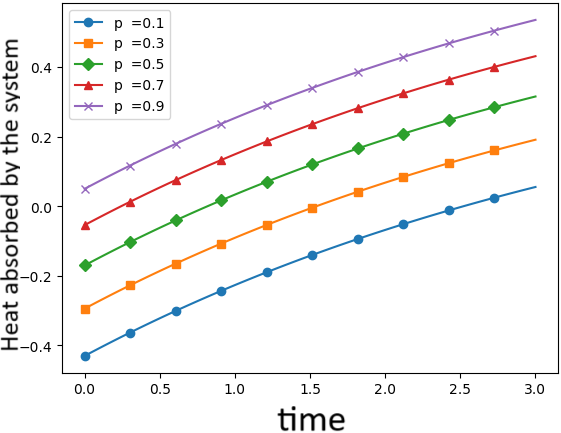} \vfill  $\left(c\right)$
 			\end{minipage}}}

 			{{\begin{minipage}[b]{.33\linewidth}
 						\centering
 						\includegraphics[scale=0.45]{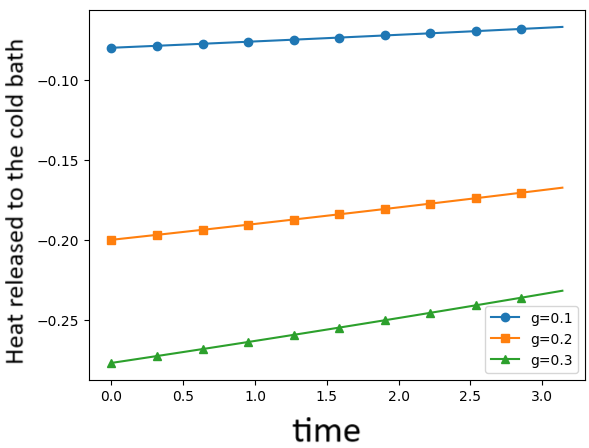} \vfill $\left(d\right)$
 					\end{minipage} \hfill
 					\begin{minipage}[b]{.33\linewidth}
 						\centering
 						\includegraphics[scale=0.45]{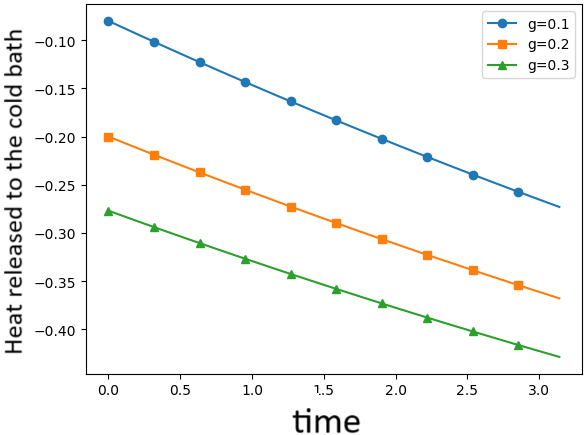} \vfill  $\left(e\right)$
 					\end{minipage} \hfill
 					\begin{minipage}[b]{.33\linewidth}
 						\centering
 						\includegraphics[scale=0.45]{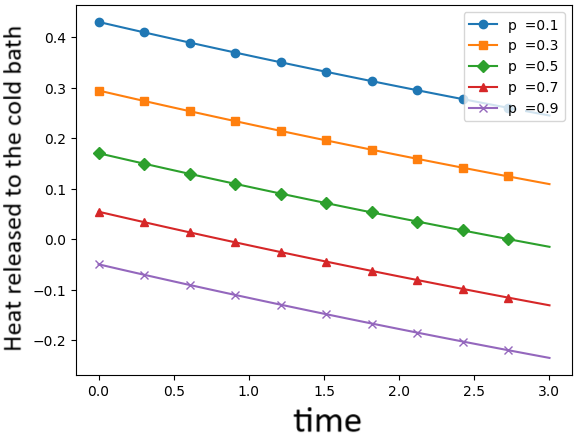} \vfill  $\left(f\right)$
 			\end{minipage}}}
 		 \caption{Dynamics of heat exchange in a two-qubit system. Panels (a) and (b) show the heat absorbed by the system, $Q_h$, in the local master equation (LME) and global master equation (GME), respectively. Panels (d) and (e) depict the heat released to the cold bath, $Q_c$, for different coupling strengths, $g$, while keeping p=1 fixed. Panels (c) and (f) illustrate the dependence of $Q_h$ and $Q_c$ on the initial state probabilities in the GME, with fixed parameters: $\omega_{A}=1$, $\omega_b=0.4$, $T_h=70$, $T_c=15$.}
                    \label{Fig7}
 		\end{figure}

 		\begin{figure}[hbtp]
 			{{\begin{minipage}[b]{.33\linewidth}
 						\centering
 						\includegraphics[scale=0.3]{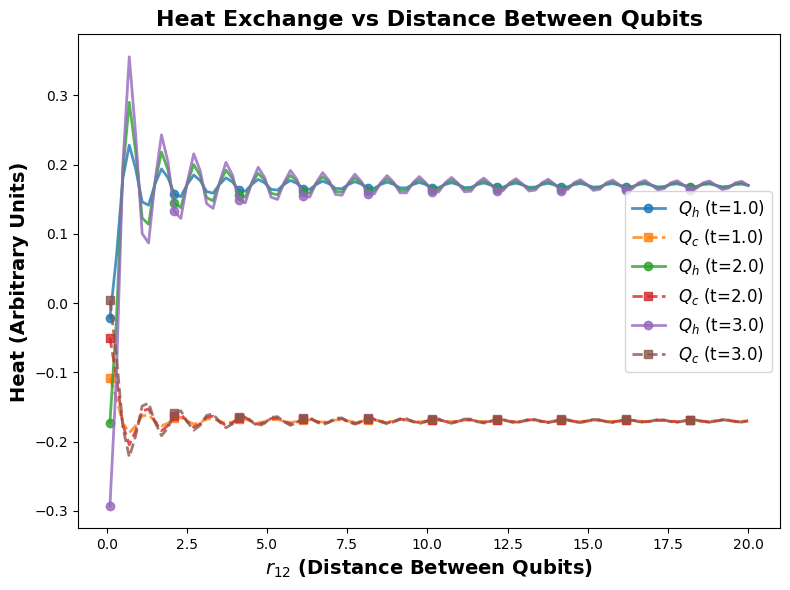} \vfill $\left(a\right)$
 					\end{minipage} \hfill
 					\begin{minipage}[b]{.33\linewidth}
 						\centering
 						\includegraphics[scale=0.3]{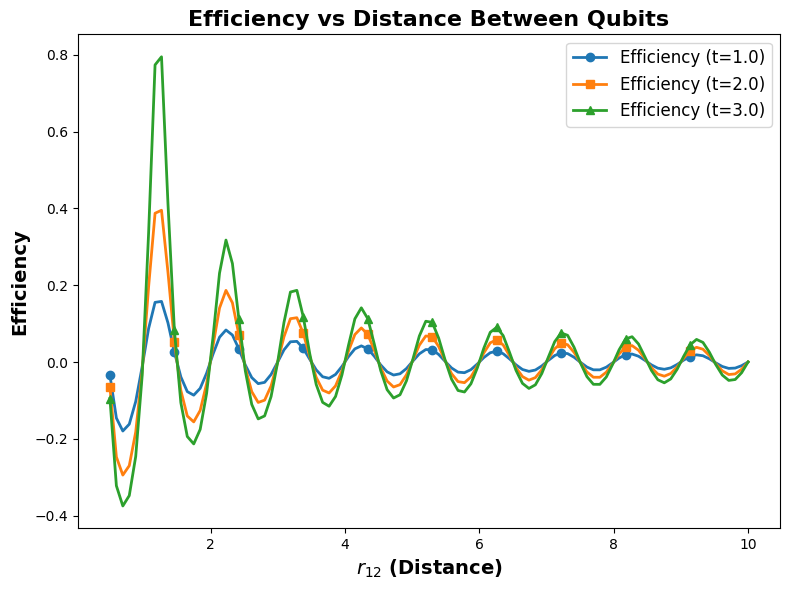} \vfill  $\left(b\right)$
 					\end{minipage} \hfill
 					\begin{minipage}[b]{.33\linewidth}
 						\centering
 						\includegraphics[scale=0.3]{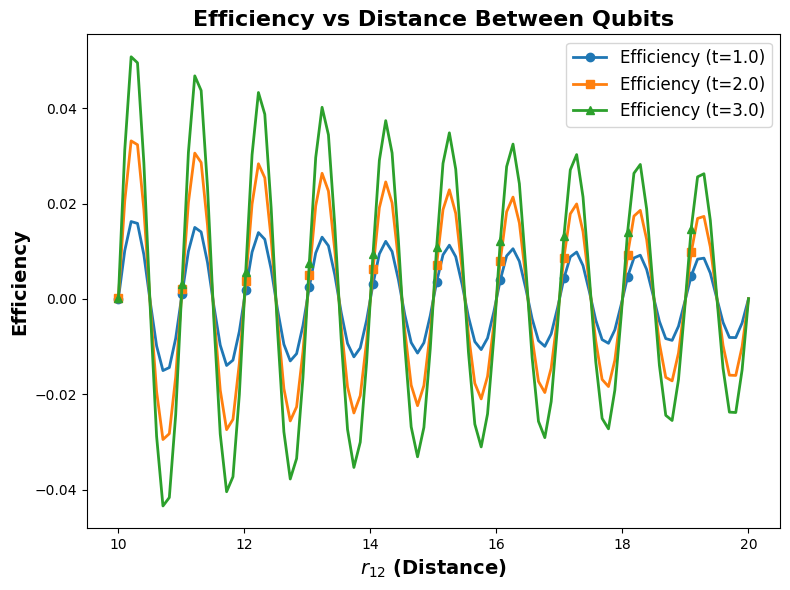} \vfill  $\left(c\right)$
 			\end{minipage}}}

 			{{\begin{minipage}[b]{.33\linewidth}
 						\centering
 						\includegraphics[scale=0.3]{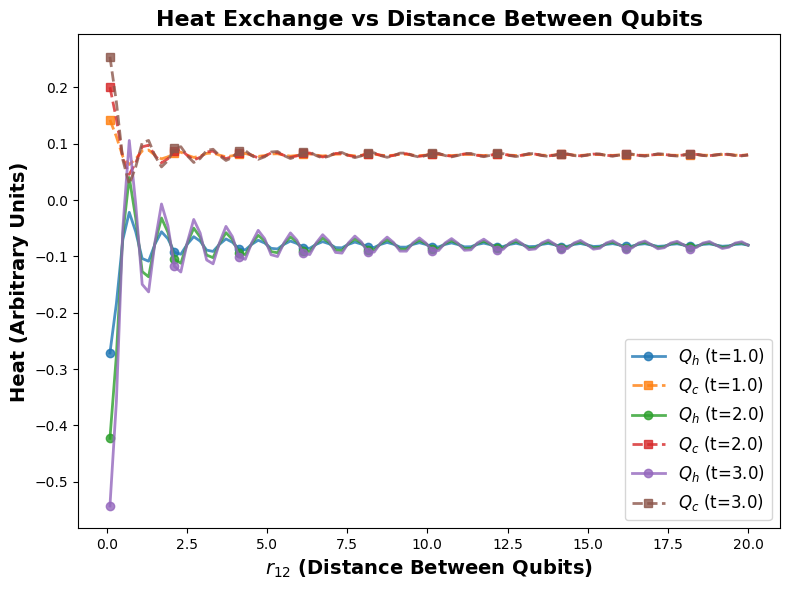} \vfill $\left(d\right)$
 					\end{minipage} \hfill
 					\begin{minipage}[b]{.33\linewidth}
 						\centering
 						\includegraphics[scale=0.3]{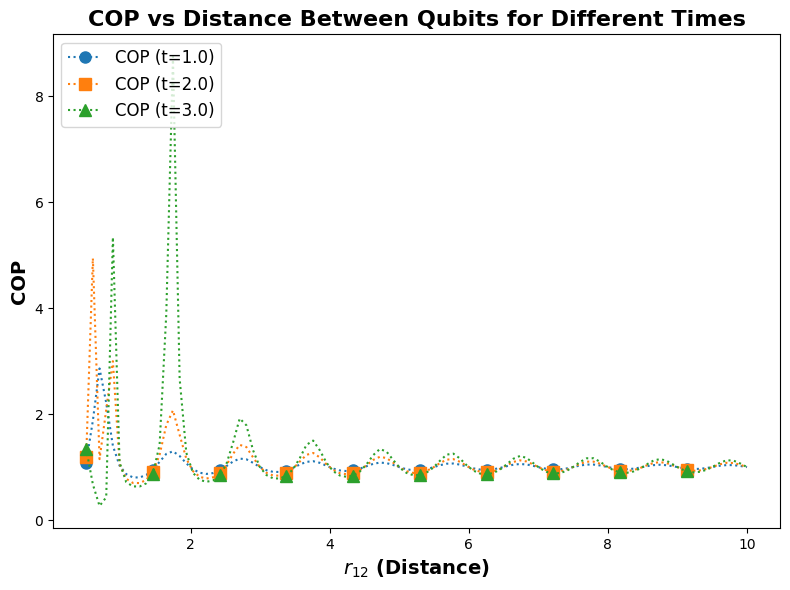} \vfill  $\left(e\right)$
 					\end{minipage} \hfill
 					\begin{minipage}[b]{.33\linewidth}
 						\centering
 						\includegraphics[scale=0.3]{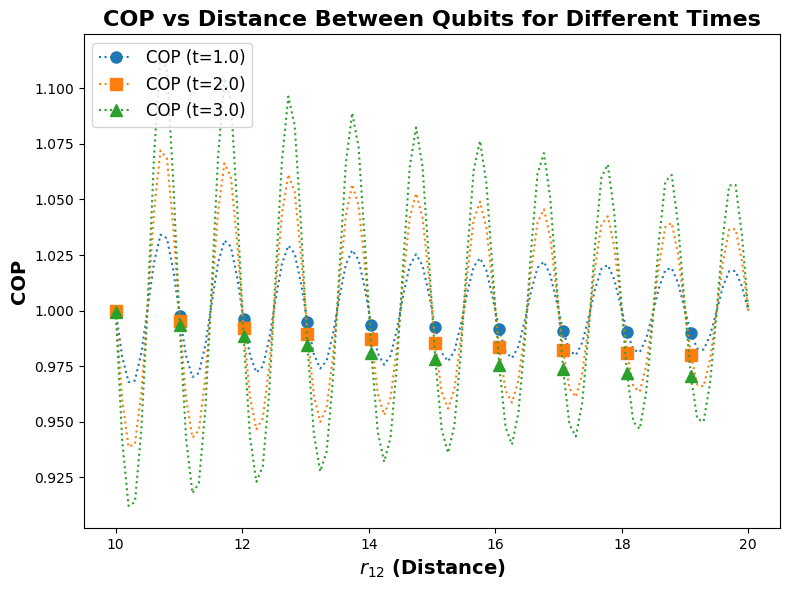} \vfill  $\left(f\right)$
 			\end{minipage}}}
 		 \caption{Dynamics of heat $Q_c$ and $Q_h$ exchange in a two-qubit system under the changed values of $r_{12}$, (a) Engine mode while (d) fridge mode,(b, c) The Efficiency of the engine in the function of $r_{12}$,(e, f) is the coefficient of performance of the fridge in the function of $r_{12}$. }
                    \label{Fig12}
 		\end{figure}
\end{widetext}

\end{document}